\documentclass[pdflatex,sn-basic]{sn-jnl}

\jyear{2021}%

\theoremstyle{thmstyleone}%
\theoremstyle{thmstyletwo}%

\theoremstyle{thmstylethree}%

\begin{document}

\title[Life and Science of Eugene Parker]{``Gene'': A personal
tribute to the Life and Science of Eugene Newman Parker}


\author[1,2]{\fnm{Arnab} \sur{Rai Choudhuri}}\email{arnab@iisc.ac.in}

\affil[1]{\orgdiv{Department of Physics}, \orgname{Indian Institute of Science}, \city{Bengaluru}, \postcode{560012}, \country{India}}

\affil[2]{\orgname{Max Planck Institute for the 
History of Science}, \orgaddress{\street{Boltzmannstrasse 22}, \city{Berlin}, \postcode{14195}, \country{Germany}}}

\abstract{This review provides a brief account of the life of Eugene Parker (1927--2022) and discusses his
contributions to plasma astrophysics. Growing up in Michigan, he went to graduate school
at Caltech and then worked at the University of Utah before shifting to the University of
Chicago, where he spent the rest of his illustrious career. Parker's most important scientific
works are discussed in the context of the historical development of plasma astrophysics. In the
study of the Sun, he made enormous contributions both to the MHD of the solar convection zone
(including the formulation of turbulent dynamo theory) and to the understanding of the outer
solar atmosphere (including the theory of coronal heating and the prediction of the solar wind).
Parker's non-solar contributions include the Parker instability in the interstellar gas and the
Parker limit of magnetic monopoles.  We also try to convey an idea of Parker's highly individualistic
personality and his very unique way of doing science. }


\maketitle

\section{Introduction}\label{sec1}
Eugene Newman Parker---known simply as ``Gene'' to his friends and admirers---was arguably the most influential scientist in the field of plasma astrophysics.  His passing away on 15 March 2022 at the age of 94 indeed marks the end of an era when research in theoretical plasma 
astrophysics could be done with a certain elegance of style---at least by the masters of the 
subject. It was a style of performing elegant
analytical calculations with simple models,
formulated on the basis of deep physical
insight, that would have far-reaching
consequences in understanding important
phenomena of nature.
That style is perhaps possible in a subject only when its foundations are being
laid down.  It is disappearing in the field
of plasma astrophysics with changing times as the subject is becoming more mature and technical.

Parker's creative career coincided with what I would call the heroic age of plasma astrophysics.
When he started research in the early 1950s, the only astrophysical locations known to have
magnetic fields were sunspots.  It was during Parker's long career that different observational
techniques established the ubiquitous presence of magnetic fields virtually everywhere in the astrophysical
universe.  On the theoretical front also, some of the first basic ideas of MHD had just been developed by
Alfv\'en, Cowling, Elsasser, Chandrasekhar, Spitzer and a few others when Parker entered the
field.  It was expected that MHD would provide the key to understanding various cosmical phenomena,
as indicated by the title {\em Cosmical Electrodynamics} of the classic monograph by \citet{Alfven1950}.
But it was still an expectation rather than a reality. Since the Sun happens to be a nearby (by
astronomical standards!) large plasma
body in which various MHD processes could be observed in detail, a major part of Parker's research
output dealt with MHD processes in the Sun.  However, Parker was very particular that he should
not be regarded only as a solar physicist, but rather as an astrophysicist. 
When I was the Executive Editor of {\em Research in Astronomy and Astrophysics} and persuaded 
Parker to write a charming scientific reminiscence \citep{Parker2014}, he titled it ``Reminiscing 
my sixty year 
pursuit of the physics of the Sun and the Galaxy''.  Parker was a humble and self-effacing man 
who rarely told people about his achievements. To the best of my knowledge, this is one of the only two
accounts of his life and career in his own words that we have. 
The other one is the lecture he gave while receiving the Kyoto
Prize, of which the transcript is available at the website:

\begin{quote}
https://www.kyotoprize.org/wp-content/uploads/2019/07/2003\_B.pdf
\end{quote}
We also refer to a journalistic account of Parker's life by M.\ 
Kaufman:
\begin{quote}
https://manyworlds.space/2022/04/12/nature-has-become-more-beautiful-physicist-eugene-parker-and-his-life-unlocking-secrets-of-the-sun/
\end{quote}
and an obituary by K.\ Tsinganos:
\begin{quote}
https://baas.aas.org/pub/2022i039/release/1
\end{quote}

The aim of the present review is to give an account of Parker's major scientific achievements along
with a few words about his life and personality. I had the privilege of being his PhD student during 1981--85. However,
as I started preparing this review, I realized that being Parker's PhD student does not automatically 
qualify somebody to write about his science.  He made such wide-ranging and varied contributions 
in different aspects of plasma astrophysics that it is impossible for one person to fully understand 
the significance of all of Parker's works at a technical level, unless that person also happens 
to be almost as brilliant as Parker himself! 

Rather than providing a catalogue of many of Parker's
works, I shall focus on a few of his outstanding contributions.  I shall try to give an idea of the historical
contexts in which these works appeared and also describe how they influenced the subsequent
development of the field.  My aim will be to
present a discussion of Parker's science in
such a manner that it is accessible to any
professional physicist rather than to experts
of plasma astrophysics alone. 
When I discuss Parker's works on those topics of which my own knowledge is
limited, my discussions will have to be
somewhat superficial. It often happens in the case of creative geniuses
in different spheres of human creativity (literature, art, music, science) that their most famous
works eclipse their other almost equally important works.  This happened to some extent in the case of
Gene Parker.  It is perhaps indisputable that the prediction of the solar wind was his most 
important work \citep{Parker58}.  However, some of his other works such as the formulation of the turbulent dynamo theory \citep{Parker55b} and the
discovery of the non-equilibrium of magnetic topologies in stellar coronae \citep{Parker72} can hardly be considered less
significant.  At the time of writing this review, Web of Science lists about 2750 citations to the
paper predicting the solar wind \citep{Parker58}, whereas there are about 1620 citations to the paper
laying down the foundations of dynamo theory \citep{Parker55b}.
A discussion of Parker's major contributions hopefully will give an idea of the breadth of his contributions. Since he had worked on
almost all important aspects of solar physics 
and many important aspects of non-solar plasma
astrophysics, this review provides a
historical account of the growth of plasma
astrophysics during 1950--1995 when Parker was
active in research.

\begin{figure}
    \centering
\includegraphics[width= 0.7\textwidth]{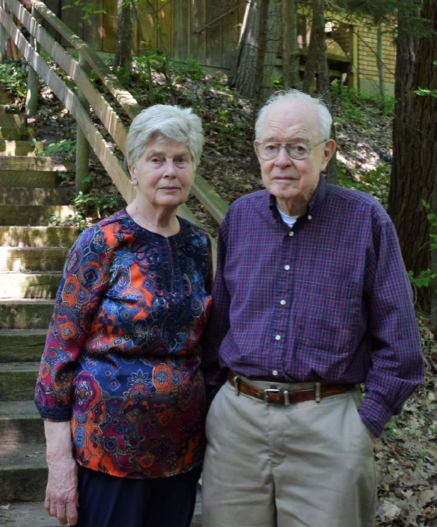}
    \caption{Gene Parker with wife Niesje in front of
    their vacation home in Michigan on the day Gene turned 90. 
    Credit: Eric Parker.}
\end{figure}

I shall try to give an idea of Parker's attitude towards science and the scientific community in a later
section (section~\ref{sec8}) of the review. However, it may be worthwhile to say a few words about Parker's style of
science at the very beginning before we enter into a detailed discussion of his scientific contributions.
What strikes one even from a superficial perusal of Parker's works is the high degree of individualism.
I have rarely come across another scientist who was so little swayed by the scientific fashions of the
day and had the courage to follow his own uncharted path of scientific investigations.  Although scientific
collaborations and multi-author papers were becoming the norm towards the later part of Parker's career,
he mostly worked on his own.  He also expected his students and postdocs to work on their own and to write
single-author papers.  He once told me that he never agreed to be a co-author in a paper unless he
himself had repeated all the calculations in the paper on his own!  Even though numerical simulations started
becoming more common with the easy accessibility of computers and even though Parker always admitted the importance
of numerical simulations, he himself never touched a computer for his research and fully 
depended only on the insights gained from analytical
calculations. A perusal of the acknowledgments of his various papers is particularly revealing.  Parker was
strongly guided by observations in his theoretical work and regularly 
discussed with observers about their new findings. Although he
often thanked observers for useful discussion, he rarely thanked another theorist in the acknowledgments 
of his papers---especially in the later part of his scientific career. Parker's papers are always marvels
of scientific composition and bear the stamp of a scientific autocrat who enjoyed doing science in 
his own terms.  He would always pay particular
attention to the logical structure of the paper.  Since Parker often dealt with complex ideas years before
others paid attention to them, it may not always be easy to read his papers.  But a reader with the prerequisite
technical knowledge can always follow the clear thread of scientific logic.  Nothing would be fuzzy or obscure.

The next section will provide a brief sketch of Parker's early life and formative years.  Then four sections will be devoted to Parker's contributions in his main research field of solar physics.
In section~\ref{sec3}, I shall make some comments about the general nature of Parker's works in solar physics and how
they transformed the field.  Then section~\ref{sec4} will discuss Parker's main contributions to the MHD of the solar
convection zone (dynamo theory and magnetic buoyancy), whereas section~\ref{sec5} will highlight his contributions to
our understanding of the outer atmosphere of the Sun (coronal heating and solar wind).  Parker's various
other important works related to the Sun will be briefly summarized in section~\ref{sec6}. After these four sections, 
section~\ref{sec7} will be devoted to 
Parker's seminal non-solar works---mostly connected with the galactic magnetic field. In section~\ref{sec8}, I shall
try to present a pen portrait of Gene Parker as a scientist and as a man. Finally, I shall end with a
few concluding remarks in section~\ref{sec9}.  

\section{The formative years and the don at Chicago}\label{sec2}

Gene Parker was born on 10 June 1927 in the small town of Houghton in Michigan with a current 
population of about 8000. One can imagine that it was really a small town where his 
childhood years coincided with the Great Depression, which impacted American small towns 
in a manner depicted in the novels of John Steinbeck. The family, however, did not stay
in Houghton, a copper mining town, for long. Gene's father Glenn Parker,
who initially worked as a mines surveyor there, shifted to other jobs and eventually started
working as an engineer for Chrysler, staying in the suburb of
Detroit.  There was a railroad yard nearby. Gene
as a small boy was fascinated by the steam locomotive
and wondered about its working principles
\citep{Parker2014}. Gene's mother Helen, who
had a BS degree in mathematics from Stanford
University, did not pursue a career and raised
three children, Gene being the eldest of the
three.  Some years later, when Glenn
Parker had retired from Chrysler, he and Helen 
moved to the warmer Arkansas, where they developed
a farm to raise cattle and chickens. There
was no telephone in this farm in the 1950s.
For many years, Gene Parker would write letters
to his parents regularly.  Luckily, the family
has preserved these letters, which give a 
fascinating glimpse of the mind of Gene
during some of the most important and creative
years of his career.

Even in his boyhood, Gene displayed
the rugged individualism of the American pioneer spirit.  When he was barely 16, he
used his earnings from summer jobs to buy a 40-acre wooded land in a remote location at
the price of \$120. He and his younger brother built a log cabin there over the course of the
next three summers---in the tradition of
Thoreau building his log cabin near the Walden pond. Gene would have to ride a bicycle
for 300 miles to go to his log cabin. After he passed away many years later and his body
was cremated, half of the ashes had been buried in the wooded land near that cabin which still has no electricity
or running water.

The USA in the days of Gene's youth was not yet the highly networked country which 
it became later. Long distance telephone calls would cost exorbitantly. People would 
travel across the country in buses and trains, since air travel was very expensive.    
Students went to colleges near their homes.  Gene went for his BS to Michigan State 
University with a tuition scholarship. Although it was not a high-ranking university, 
Gene was lucky to have some extremely dedicated physics teachers who urged Gene to go 
for a top graduate school.  Being not from a very well-off family, Gene had to save 
some money by working for six months as technician at the Chrysler lab in Detroit and 
then took a 72-hour Greyhound bus ride to Caltech, where he was a graduate student 
during 1948-51. Caltech did not give him any financial aid because people there did 
not know how to calibrate a grade report from a university in Michigan! Gene found 
Pasadena to be rather expensive and realized that his savings would run out in a 
few months.  Luckily, he had done well in the quantum mechanics course taught by 
William Fowler, who later won the Nobel Prize for his work in nuclear astrophysics.  
When Fowler came to know that Gene was without financial aid, he immediately telephoned 
the Dean and insisted that this boy must be given a teaching assistantship \citep{Parker2014}.  Gene 
survived in Caltech with that.  His supervisor was H.\ P.\ Robertson, known for 
the Robertson-Walker metric in cosmology, who was working on other things at that 
time.  He urged Gene to work out the theory of some structures observed in the 
interstellar medium. Gene's other mentor at Caltech was Leverett Davis, who 
interestingly was to work on an extension of Gene's model of the solar wind
many years later \citep{WD67}.

\begin{figure}
    \centering
\includegraphics[width= 0.95\textwidth, trim={3cm 19cm 1cm 0}, clip]{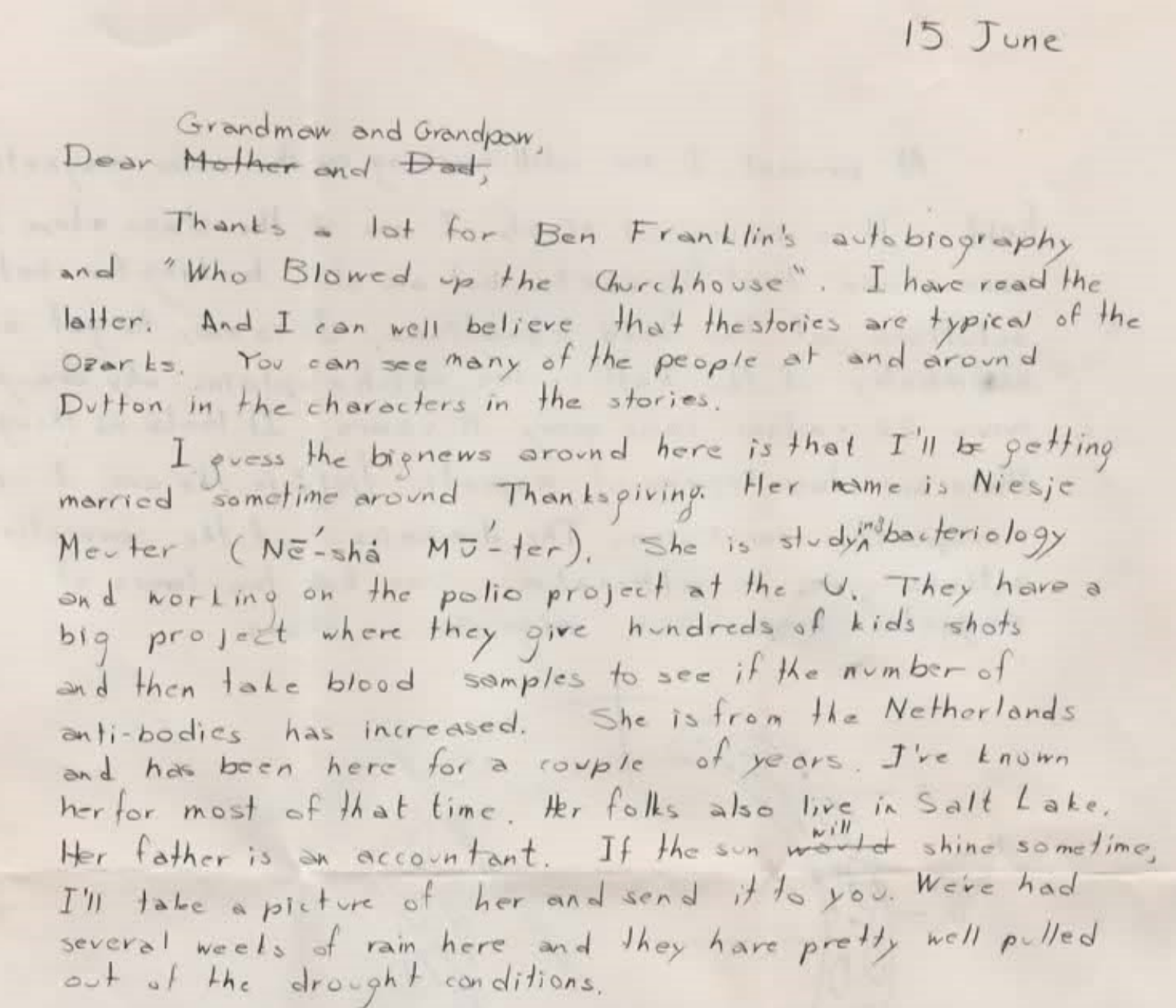}
    \caption{The beginning of Gene's letter to his parents dated 15 June 1954
    informing them of his upcoming marriage in an
    unceremonious manner. This letter was written soon after Gene came to
    know that his younger brother had a son so that his parents had become
    grandparents. Credit: Eric Parker and Susan Kane-Parker.}
\end{figure}

After PhD, Gene worked for a few years at the University of Utah – first as 
instructor and then as research associate of Walter Elsasser.  Gene always had a 
tremendous respect for Elsasser and regarded Elsasser as his real mentor 
who introduced him to the problem of astrophysical dynamos.  Although Gene 
did some pathbreaking works in Utah, the authorities there felt that he 
was not doing interesting enough research and did not want to give him tenure. At 
that time, John Simpson was building state-of-the-art instruments at the 
University of Chicago to study cosmic rays and wanted to hire a theorist 
who could help them in selecting the right scientific questions to study. 
Chandrasekhar, Simpson's colleague at Chicago, suggested the name of Gene 
to Simpson.  Gene came to Chicago in 1955 and remained there all his life, 
rising through the academic ranks and retiring in 1995.  Gene served
as Chairman of the Department of Physics during 1970--72 and as Chairman of
the Department of Astronomy and Astrophysics during 1972--78. He was also
the Chairman of the Astronomy Section of the National Academy of Sciences
during 1983--86.

While at Utah, Gene met Niesje, who was to become his wife and life partner.
Niesje grew up in the Netherlands, where her family lived through the difficult years of
the Nazi occupation and then immigrated to Utah after World War II. With an initial
training in bacteriology, she eventually got a job at the University of Chicago 
Graduate School of Business, rising to the position of Associate Director of Computing 
Services there.  Gene and Niesje were married
in Salt Lake City on November 24, 1954
shortly before they moved to Chicago.
Their children---Joyce and Eric---were born in Chicago.  Figure~2 shows a part of Gene's
letter to his parents informing them of his
impending marriage in an unceremonious way.

For many years, the family
lived in a house at Homewood in the suburb of Chicago.  When there would be an academic
visitor whom Gene particularly admired, he would invite the visitor for dinner to his
home with members of his group.  I remember going for dinner to their home when
Nigel Weiss and Henk Spruit visited. Many years later, when I visited Chicago from
India, I had the honour of being the guest for whom Gene gave a dinner party.  
In some ways, Gene was the traditional husband who left the job of preparing the
dinner to Niesje, an outstanding cook.  However, Gene was the gracious host
who would set the table and serve the guests.  Gene himself never drank alcohol.
I may mention that, when Gene's graduate student Tom Bogdan, a year senior to me,
graduated, I gave a dinner party in my student apartment and invited Gene.  Tom was
a little alarmed that I had the temerity of inviting Gene to a typical grad student
apartment, which was not in particularly great shape.  But Gene and Niesje came and enjoyed themselves. Figure~1 shows Gene
and Niesje taken on Gene's 90th birthday.

\begin{figure}
    \centering
\includegraphics[width= 0.95\textwidth]{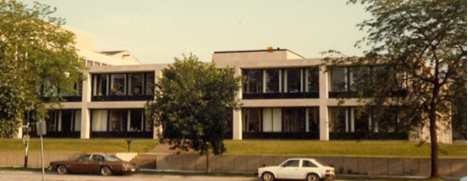}
    \caption{The Laboratory for Astrophysics
    and Space Research (LASR) in the University of
    Chicago campus, where Parker had his office
    for several years.}
\end{figure}  

For several decades, Gene worked in the Laboratory for Astrophysics and Space Research (LASR),
a building of modest size on the University of Chicago campus which housed the
cosmic ray research group headed by John Simpson. However, two of the most beautiful corner
offices in that building were occupied by two theorists: Chandrasekhar and Parker. 
In Figure~3 showing LASR, the upper left corner room was Gene's office and the upper
right corner room was Chandra's office.  It will probably be difficult to identify another building
anywhere in the world from which so many outstanding contributions to plasma astrophysics
originated. This building LASR was considerably 
restructured a few years ago. After Chandra's
passing away, Gene wrote a review describing Chandra's contributions to MHD \citep{Parker96}.
Figure~4 shows photographs of Gene
at two stages of life---as a university student and as a university professor.  The photograph on the right side was taken by me during my student days. I particularly like this photograph, because looking at Parker standing in front of his huge collection of books and 
journals in his office, you get a feeling about the kind of scholarly scientist that he was.  
Gene's office was always in complete disorder---with piles of papers and books strewn 
all around. I often wondered how he managed to find anything in his office. Although Chandra 
and Parker were mathematical physicists of somewhat similar mould and were good friends, 
their personalities were totally different.  Everything in Chandra's office would be 
perfectly ordered: every piece of paper in its correct place!

\begin{figure}
    \centering
\includegraphics[width= 1.0\textwidth]{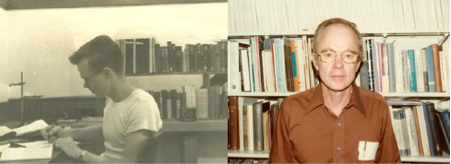}
    \caption{Gene Parker at different stages of
    his career: (i) As a graduate student at Caltech,
    and (ii) As a don at the University of Chicago.}
\end{figure}

A teetotaller, Gene was a man of very simple and almost austere habits. 
He would usually commute by
the Illinois Central train from his home in Homewood
to the University of Chicago campus, although the train station was not exactly next
door to the building LASR where we worked. 
Though he knew quite a bit about cars, 
he would drive only if he
needed the car for something during the day.
However, he allowed himself to indulge in one
luxury.  He always liked having personal
copies of the books and journals he would look at for his academic work rather than consulting
them in the library. In those pre-internet 
days when one would normally go to the library
to look up journals, Gene had a personal
collection of several journals, such as
{\em Astrophysical Journal, Solar Physics, 
Geophysical and Astrophysical Fluid Dynamics}.
One wall of his office was completely converted
into a bookshelf where his books and journals
were kept.  One can get a glimpse of this personal
library in the right side photograph of 
Figure~4.

After this account of Gene Parker's early
life and scientific career, we shall now 
turn to his science in sections~3--7.  I 
shall try to give a pen portrait of Gene
as a scientist and a member of the scientific
community in section~8, since such a pen
portrait may be better appreciated by readers
after knowing about his science.

\section{Gene Parker and the growth of solar physics (and beyond)}\label{sec3}

Before describing the specific details of Gene Parker's works, I would first like to make a
few overall comments about the way Parker transformed our theoretical understanding of
various aspects of solar activity.  A branch of science in which different important 
topics are interconnected through some unifying principles always possesses a special
kind of intellectual appeal. Parker's monumental contribution to solar physics was to gradually
build up over the years a grand edifice in which we can see the connections among the
different aspects of solar activity, through the common thread of the magnetic field and
its different manifestations in the solar plasma.

Instead of discussing Parker's works chronologically, let me highlight some of Parker's
key contributions in a logical order best suited to illuminate the unified structure of
the field solar MHD. Details of these works will be given in sections 4 and 5.
The first central question is how magnetic fields arise in the Sun
and other astrophysical bodies.  Quite early in his career, \citet{Parker55b} formulated turbulent
dynamo theory which provides an understanding of how magnetic fields may arise in regions of
convection inside a rotating astrophysical body.  \citet{Parker55b} also applied the basic ideas of dynamo
theory to work out an ingenious model of the 11-year sunspot cycle. While today we may not agree
with Parker's model of the 11-year sunspot cycle in all its details, it is indisputable that
Parker's 1955 paper \citep{Parker55b} is the most influential paper in the history of dynamo theory and is still
the starting point for anybody wanting to understand how magnetic fields are generated in
astrophysical systems. The magnetic fields generated in the solar convection zone have to come
to the solar surface and into the atmosphere.  In the same year 1955, in which Parker formulated dynamo
theory, he also wrote the fundamental paper on magnetic buoyancy to explain why magnetic fields
of the Sun (or rather parts of them) emerge from the Sun's interior through its surface \citep{Parker55a}.  The magnetic fields
which emerge through the solar surface are the cause of various activities in the corona.
When the theory of magnetic reconnection in the corona was being developed, Parker wrote a
key paper giving rise to the idea of the so-called Sweet--Parker reconnection rate \citep{Parker57}.
A few years later, \citet{Parker72} discovered that coronal magnetic fields, while trying to relax
to configurations of magnetostatic equilibrium, tend to produce many current sheets
(i.e.\ regions of magnetic reconnection) in the corona, within which heat is produced
by the conversion of magnetic energy.  While this mechanism discovered
by Parker may not be the sole mechanism for generating heat in the corona (magnetoacoustic
waves dissipating in the corona may also contribute in a parallel mechanism), Parker's 
theory appears applicable for explaining the heating of the coronal magnetic loops, the
hottest regions in the lower corona.  Finally, the crowning achievement of Parker's illustrious
career was to show that the hot corona would drive
an outward plasma flow which he named 
the solar wind \citep{Parker58}.  
This radical prediction was at first viewed in the community with
scepticism and perhaps even disbelief, until support
for it came from space observations within a few years.
Remarkably,
the prediction of the solar wind, which is caused by
the hot corona,
came several years before there was much understanding
of what heats the corona.  

\begin{figure}
    \centering
\includegraphics[width= 1.0\textwidth]{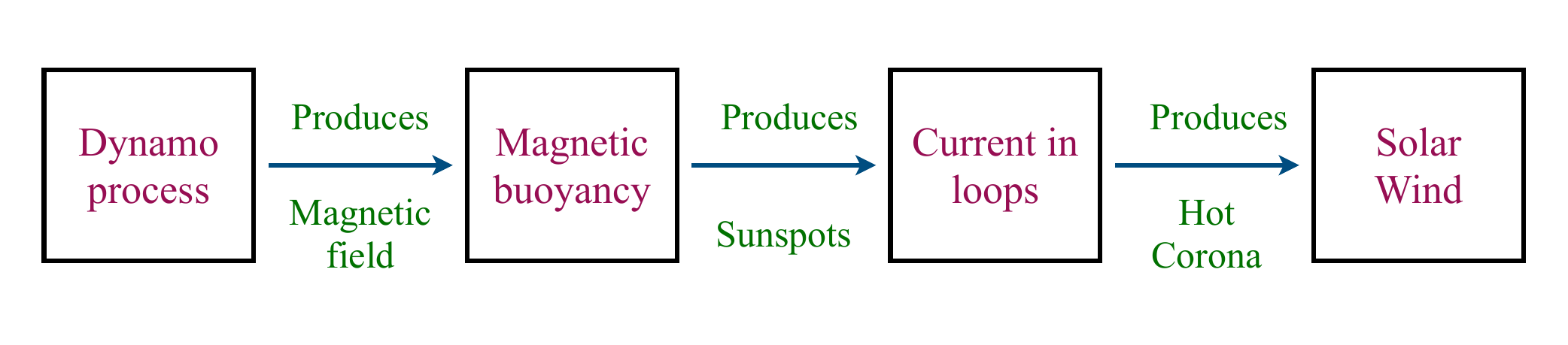}
    \caption{The chain of causes-and-effects connecting different
    solar phenomena.}
\end{figure}
  
Why do many of us consider Gene Parker to be the greatest solar physicist of our 
time---or perhaps of all time? To understand how he transformed the field, we may look 
at two books published at the beginning and at the end of the most creative phase 
of Parker's career \citep{Kuiper53,Priest82}.  The chapter on solar activity by \citet{Kiepen53} in the first 
book gives various kinds of data about different solar phenomena without any clue 
how to unify them.  The chapter on solar MHD by \citet{Cowling53} summarized the basic principles 
of MHD and expressed the hope that in future they might be useful for studying the
Sun.  Apart from a discussion of a long-discarded model of sunspots due to 
Alfv\'en, the chapter presented almost no real applications to the Sun.  However, when we
look at Priest's book published in 1982 \citep{Priest82}, we realize that the subject is already organized 
and connected essentially in the way we would do today.  Solar physics had become 
a unified science in the intervening three decades during which
Gene Parker formulated dynamo theory, gave the idea of magnetic buoyancy, developed 
the theory of coronal heating and predicted the solar wind. It will be difficult 
find another similar example in any branch of astrophysics of one individual striding 
over the field as a colossus like this. I show in Figure~5 a cartoon of how the various
topics in the study of solar MHD are logically connected to each other through a chain
of causes-and-effects.  What is most amazing is that Parker almost
single-handedly discovered virtually all the important links needed for interconnecting
various aspects of solar activity, as should be clear from our discussion. A historical account of Parker's works
on solar MHD essentially becomes a history of the
field of solar MHD during its most crucial years
of development!

The next two sections will be devoted to Parker's contributions in the MHD of the 
solar convection zone and the solar corona.  Apart from establishing the interconnecting
network sketched in Figure~5, Parker also explored several side lanes of solar physics
to develop theoretical ideas to explain many other solar phenomena.  Some of Parker's
other important contributions to solar physics will be discussed in section~6 before we turn
to his non-solar contributions in section~7.

\def\Bb{{\bf B}}
\def\vb{{\bf v}}
\def\pa{\partial}
\def\bb{{\bf b}}

It may be mentioned that,
during the early years of Gene Parker's career, not much was
known about stellar activity.  During the last few decades,
it has been realized that many solar-like stars have starspots much larger
than sunspots, stellar flares much more powerful than solar flares
and activity cycles similar to the Sun: see the review by
\citet{Chou17}.  It is now clear that Parker's works on the Sun
have a much broader significance, providing us the framework for
understanding many aspects of stellar activity.

Since the majority of Parker's papers (not all of
them!) had been based on the macroscopic equations
of MHD, we provide a quick recapitulation of these
equations before we start a detailed discussion of Parker's
works. What Parker could coax out of these simple-looking
equations almost appears magical to us today.
The basic idea of MHD is that the fluid
velocity $\vb$ and the magnetic field $\Bb$ act
on each other.  Since Parker mainly used Gaussian
units in his publications, we now write down the
MHD equations for $\vb$ and $\Bb$ in Gaussian units. 
The time evolution equation of
$\vb$ is
$$\frac{\pa \vb}{\pa t} + (\vb. \nabla) \vb = 
- \frac{1}{\rho} \nabla \! \left(p+ \frac{B^2}{8 \pi}\right) 
+ \frac{(\Bb. \nabla)\Bb}{4 \pi \rho} + {\bf F}, \eqno(1)$$
where $\rho$ is the density, $p$ is the pressure and
${\bf F}$ is a body force per
unit mass such as gravity: see, for example, \citet{Chou98}, sect.\ 14.1.
When we leave out the magnetic terms, which arise
from the Lorentz force, we are left with the 
well-known Euler equation of fluid mechanics.  It is
not difficult to explain the physical significance
of the magnetic terms. The term $B^2/8 \pi$, which
appears with the gas pressure $p$ inside the expression
of a gradient force, is clearly of the nature of pressure.
We call it {\em magnetic pressure}.  The other magnetic force
term involving $(\Bb. \nabla) \Bb$ would be zero when
the magnetic field lines are straight.  It arises only when
the magnetic field lines are bent and tries to straighten
them.  It is called {\em magnetic tension}. To understand
how $\vb$ acts on $\Bb$, we now have to look at the time
evolution equation of the magnetic field, which is known
as the {\em induction equation}:
$$\frac{\pa \Bb}{\pa t} = \nabla \times (\vb \times \Bb) + 
\eta \nabla^2 \Bb, \eqno(2)$$
with
$$\eta = \frac{c^2}{4 \pi \sigma}, \eqno(3)$$
where $c$ is the speed of light and $\sigma$ is the 
electrical conductivity:
see, for example, \citet{Chou98}, sect.\ 14.1.  Since $1/\sigma$ is
the resistivity, the last term in (2) corresponds to
the decay of the magnetic field due to the resistivity
of the medium.  The term $\nabla \times (\vb \times \Bb)$
can be shown to imply that the velocity field $\vb$ makes
the magnetic field move with the plasma---an effect first
discovered by \citet{Alfven43} and often referred to as {\em
Alfv\'en's theorem of flux freezing}.

\section{MHD of the solar convection zone}\label{sec4}

The heat generated by nuclear fusion near the centre
of the Sun is transported by convection from about $0.7 R_{\odot}$
to $R_{\odot}$: see, for example, \citet{Chou2010}, sects.\ 3.2.4 and
4.4.  
This region having the shape of a spherical
shell is known as the convection zone and is found to be
unstable to convection.  Although we observe magnetic fields
only on the solar surface and infer the nature of
magnetic fields within the convection zone only on the basis
of indirect arguments, reasonable theoretical considerations
suggest that the plasma $\beta$, i.e. the ratio of the gas
pressure to the magnetic pressure, is much larger than 1 
within the convection zone except 
very near the surface. This is in contrast to the corona
where the plasma $\beta$ is less than 1 in many regions and
the magnetic field controls the dynamics.  Since the magnetic
field does not control the dynamics of the convection zone,
it may a priori appear that the role of the magnetic fields
may not be so important there.  In reality, however, the
convection zone is of central interest to MHD as the region
where the magnetic field originates.

As already mentioned in section~3, Parker wrote one seminal
paper on how magnetic fields are produced by the dynamo process
in the convection zone \citep{Parker55b} and another seminal
paper on how these fields rise through the convection zone to
reach the surface due to magnetic buoyancy \citep{Parker55a}.
Both these papers were submitted to {\em Astrophysical Journal}
on October 18, 1954, indicating that Parker looked at these
related problems from a unified viewpoint, although he
eventually wrote two separate papers for two parts of the
problem.  The dynamo paper was revised on May 11, 1955, but
no such revision date is given for the buoyancy paper, suggesting
that it was probably accepted readily by the
referee without requiring
any revisions.  The buoyancy paper \citep{Parker55a} appeared
before the dynamo paper \citep{Parker55b}.  Curiously, Parker
motivates the discussion of the buoyancy paper by summarizing
some relevant results from the dynamo paper in the very first
paragraph of the Introduction, although the dynamo paper was
yet to appear! I may point out another perhaps irrelevant
fact.  Parker used SI units for electromagnetic quantities
in these two early papers,
although in later life he always used Gaussian units.  By the time
he wrote his solar wind paper three years later, he had already
switched over to Gaussian units.

We shall now discuss the buoyancy paper \citep{Parker55a} before
turning to the dynamo paper \citep{Parker55b} after that.

\subsection{Magnetic buoyancy}

Often one finds two sunspots side by side approximately at the
same latitude.  It was discovered by
\citet{Hale19} that two
sunspots in such a pair usually have opposite magnetic polarities.
The aim of \citet{Parker55a} was to provide the first satisfactory
theoretical explanation of this important observation.  It was
already known for nearly a century that the angular velocity at
the solar surface varies with latitude, becoming a little bit weaker at higher
latitudes.  Although nothing was known in 1955 about the nature of
differential rotation underneath the solar surface, the surface
observations forced the conclusion that there had to be a variation
of angular velocity in the interior. As a
consequence of flux freezing, the differential rotation in the solar interior was expected to stretch out
the magnetic field inside the
Sun to produce a
strong toroidal component (i.e.\ a component in the $\phi$ direction
in spherical coordinates with respect to the Sun's rotation
axis as the polar axis), unless magnetic field
lines were lying exactly on the contours of constant
angular velocity \citep{Ferraro37}.  If a part of the toroidal magnetic field produced due to the
stretching by differential rotation 
becomes buoyant, \citet{Parker55a} realized that this part
would rise to the surface to produce a bipolar sunspot pair, as
sketched in Figure~6 taken from Parker's paper.  \citet{Parker55a} gave
a disarmingly simple argument for how a part of the toroidal field
may become buoyant.  Consider a region of strong toroidal field
surrounded by gas without much magnetic field---a configuration
which we may call a {\em flux tube}. Inside the flux tube, we would
have magnetic pressure $B^2/8 \pi$ in addition to the gas pressure
$p_i$, which follows from (1).  This 
total pressure has to be balanced by the external gas pressure $p_e$
leading to the condition
$$ p_e = p_i + \frac{B^2}{8 \pi}.\eqno(4)$$
It follows that $p_i < p_e$, which may often imply that the internal
density of the flux tube would be less than the external density.
If this is the case, then that part of the flux tube with lower density would be buoyant
and rise against the gravitational field of the Sun to reach the
surface where it produces the bipolar sunspot pair.

\begin{figure}
    \centering
\includegraphics[width= 0.5\textwidth]{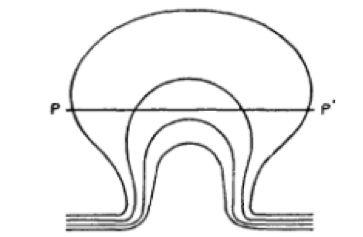}
    \caption{A sketch from the magnetic buoyancy paper \citep{Parker55a}  showing how a part of
    the magnetic field has risen to produce a bipolar
    sunspot pair at the solar surface. Reproduced by permission of the AAS.}
\end{figure}

In order to produce a bipolar sunspot pair, only a part of the toroidal
magnetic field should rise and the two ends of this part should remain
`clamped'.  \citet{Parker55a} did not present much discussion of any possible 
clamping mechanism in the original paper on magnetic buoyancy, except
to mention that the length $L$ of the toroidal flux tube should be
sufficiently large to ensure that magnetic tension would not be too strong
to oppose magnetic buoyancy.  Parker revisited the problem of magnetic
buoyancy two decades later \citep{Parker75}, when he presented a more
detailed analysis which provided a clue for the clamping mechanism.
\citet{Parker75} showed that magnetic buoyancy gets reinforced within
the solar convection zone where the temperature gradient is slightly
stronger than the adiabatic gradient, but the magnetic buoyancy would
be suppressed in the regions of subadiabatic temperature gradient
below the bottom of the convection zone. Suppose we have a toroidal
flux tube slightly below the bottom of the solar convection zone, of
which a part has come within the convection zone by some means.  Then that
part would become buoyant, whereas buoyancy would be suppressed in other
parts which remain anchored.  This theoretical idea readily suggests
the possibility of numerical simulations. From the full equations of
MHD, \citet{Spr81} derived an equation for the dynamics of a `thin'
flux tube, of which the radius of cross-section is much smaller than
various scale heights.  \citet{Moreno86} carried out the first
simulation  of the buoyant rise of a magnetic flux tube always lying
in a vertical plane.

Hale's collaborator Joy had noted that sunspot pairs did not appear
at exactly the same latitude, but usually had a small tilt with the
sunspot in the forward direction (with respect to the rotation axis)
lying slightly closer to the equator, and this small tilt was found to
become larger at higher latitudes \citep{Hale19}.  The existence
this tilt in bipolar sunspot pairs and its increase with latitude is
nowadays referred to as {\em Joy's law}.  One important question was whether
this tilt could arise from the effect of the Coriolis force due to the
Sun's rotation acting on the rising flux tubes.  We 
\citep{Chou89, Dsilva93} carried out the first 3D simulations of the
rise of flux tubes in a spherical geometry with the Coriolis force
included.  We found that the results of simulations matched the
observational data of Joy's law only if the magnetic field at the bottom
of the convection zone had a strength of about $10^5$ G 
\citep{Dsilva93}.  This result was soon confirmed by independent simulations
of other groups \citep{Fan93, cali95}. This provided
the first tight constraint on the value of the toroidal magnetic field
at the bottom of the convection zone and played a vital role in the
development of the solar dynamo theory, as we shall point out in
subsection~4.2.

Within the last few years, increased powers of computers have enabled a
few groups to go beyond the thin flux tube equation assumption and model the buoyant
rise of flux tubes using the full MHD equations.  Instead of entering
a discussion of this subject, we refer the readers to the excellent
review by \citet{Fan21}. 

\subsection{Solar dynamo}

A key issue in astrophysical MHD is to understand how
magnetic fields arise in astrophysical systems.  Is it
possible to have some flows in electrically conducting
fluids which would sustain a magnetic field?  This became
the central question of 
what has come to be known as {\em dynamo theory}. The early history
of this field has been summarized by \citet{Moff78},
Chapter~1.

The first important
landmark in dynamo theory was a negative theorem due to
\citet{Cowling33}, who showed that an axisymmetric flow
cannot sustain an axisymmetric magnetic field. It was
conjectured by some that {\em Cowling's theorem} may be a 
special case of a more general theorem that fluid flows
cannot sustain magnetic fields against Ohmic decay.  If
that were the case, then dynamo theory would have no
solution within the framework of MHD.  It is rumoured
that even Einstein held this view \citep{Krause93}. Those who still tried
to solve the dynamo problem knew that they had to incorporate
non-axisymmetric flows \citep{Elsasser46}.

If convection takes place in an astrophysical body which
is undergoing rotation, then the convective motions are
affected by the Coriolis force and become helical in nature.
It was the great insight of \citet{Parker55b} to realize
that such helical convective motions can sustain magnetic
fields, provided certain conditions are satisfied.  Since
these helical convective motions are turbulent, they are
certainly not axisymmetric and one easily avoids
Cowling's theorem.  \citet{Parker55b} developed a mean field
theory of averaging over turbulence and arrived at the 
famous {\em dynamo equation}, the central equation in the theory
of turbulent dynamos.  The modelling of all astrophysical
dynamos since then has been based on the fundamental ideas 
developed by Parker in this seminal paper \citep{Parker55b}.
For a pedagogical discussion of turbulent dynamo theory
and its application to the Sun, the readers are referred
to \citet{Chou15}.

\begin{figure}
    \centering
\includegraphics[width= 0.7\textwidth]{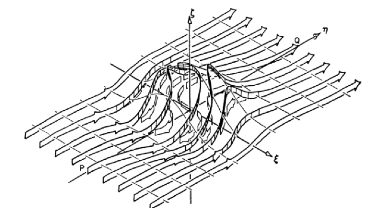}
    \caption{A figure from the dynamo paper \citep{Parker55b}  showing a few toroidal magnetic field lines to explain how the poloidal field is produced and how the dynamo
    wave for modelling the solar cycle arises. Reproduced by permission of the AAS.}
\end{figure}
\citet{Parker55b} himself solved the dynamo equation in
a situation appropriate for the Sun and obtained a wave-like
solution. Observationally, it is found that sunspots appear
at lower and lower latitudes with the progress of the solar
cycle. This was interpreted as a dynamo wave propagating
equatorward and Parker's dynamo wave solution was proposed an an
explanation of the solar cycle. It was a characteristic
of Parker's way of thinking that he always wanted to have a
physical understanding of any unusual result which he derived
mathematically.  After presenting the dynamo wave solution,
\citet{Parker55b} gave a physical explanation of how this
solution arises. This physical explanation is simply vintage
Gene Parker.  Anybody who wishes to have an understanding
of Gene Parker's unique way of thinking about physics
must read this
physical explanation of the dynamo wave in his own
words, which was reproduced more
clearly in \citet{Parker79a}, pp. 632--633.

Parker's dynamo paper \citep{Parker55b}, by far the most
influential paper in the history of this subject, had three
remarkable achievements.
\begin{enumerate}
    \item The feasibility of dynamo action within the 
    framework of MHD was demonstrated for the first time.
    \item It was shown---probably for the first time---that
    turbulence, which we normally
    expect to produce disorder, can give rise to coherent
    structures like the large-scale magnetic field.  The
    study of how coherent structures emerge out of turbulence
    due to an {\em inverse cascade} later became an important field
    of research.
    \item An attractive model of the solar cycle was proposed
    for the first time.
\end{enumerate}
Several of Parker's classic papers---notably the solar wind
paper \citep{Parker58}---provide immensely pleasurable reading.
However, the original dynamo paper was not an easy read.  Going
through that paper, one feels that \citet{Parker55b} was
struggling to put forth several difficult concepts which were
completely unfamiliar to the majority of astrophysicists in that
era. Probably very few people read or understood the paper when
it was published. According to Web of Science, the paper
received only 15 independent citations during the decade 1956--65!

After \citet{Steen66} developed a more systematic procedure
for averaging over turbulence in the dynamo problem, the subject
became more accessible.  They used the symbol $\alpha$ for the
coefficient which captures the essence of helical turbulent
motions (Parker had used the symbol $\Gamma$).  As a result,
the effect of helical turbulence twisting a magnetic field
came to be known as the {\em $\alpha$-effect}. \citet{Parker55b} had
earlier constructed a model for the solar dynamo by solving
the dynamo equation in a rectangular geometry without 
boundaries.  \citet{SK69} constructed more realistic
models of the solar dynamo by solving the dynamo equation in
a spherical geometry appropriate for the Sun.  Apart from the $\alpha$-effect,
the differential rotation of the Sun indicated by $\Omega (r, \theta)$
is the other crucial quantity responsible for dynamo action.
That is why this type of dynamo models came to be known as the
{\em $\alpha \Omega$ dynamo models}.  The necessary condition to
make the dynamo wave propagate equatorward in agreement with
observational data was found to be
$$\alpha \frac{\partial \Omega}{\partial r} < 0 \eqno(5)$$
in the northern hemisphere.  This condition is often referred
to as the {\em Parker--Yoshimura sign rule}---after \citet{Parker55b}
who obtained a primitive version of this condition from his
calculations in rectangular geometry and after \citet{Yosh75}
who generalized the condition by considering a spherical
geometry.  In the 1970s and 1980s when not much was known
about the differential rotation of the Sun below its
surface, different groups constructed models of the solar
dynamo by specifying $\alpha$ and $\Omega (r, \theta)$ in such
a manner that the condition (5) was satisfied.

Although the $\alpha \Omega$ model of the solar dynamo appeared
reasonable and attractive, some limitations of this model became
apparent in the 1980s. When {\em helioseismology}---the study of solar
oscillations---succeeded in mapping the differential rotation in
the interior of the Sun, it turned out to be very different from
what used to be assumed in the early $\alpha \Omega$ dynamo models.
Another serious difficulty arose when simulations of magnetic
buoyancy suggested that the toroidal magnetic field inside the Sun must be
as strong as $10^5$ G, as pointed out in subsection~4.1. The $\alpha$-effect
involves the twisting of the magnetic field by helical fluid motions.
For such twisting to take place, the magnetic field has to be 
weaker than about $10^4$ G. The magnetic field strength of $10^5$ G
suggested that the $\alpha$-effect would be suppressed inside the
Sun.  The older $\alpha \Omega$ dynamo models had to be modified and
amended in important ways.  Parker himself became aware of some of
these difficulties and proposed in a later paper \citep{Parker93} that
one way of circumventing them may be to develop a model
in which the differential rotation and the $\alpha$-effect operate
in different layers.

An alternative to the $\alpha$-effect was suggested quite early by
\citet{Bab61} and \citet{Leighton69}. In the {\em Babcock--Leighton mechanism},
the decay of tilted bipolar sunspots gives rise to magnetic fields
similar to what one would get from the $\alpha$-effect.  It was realized
in the 1970s and 1980s that there was a fluid flow at and 
underneath the solar surface known as {\em meridional circulation}.
In a new type of dynamo model developed in the 
1990s, known as the {\em flux transport dynamo
model}, it was found that the Babcock--Leighton mechanism, when
combined with meridional circulation, produced particularly good
fits to various aspects of the solar cycle \citep{WSN91,CSD95,Durney95}.  This model has become
increasingly popular---especially after a prediction for the following
cycle based on this model \citep{CCJ07} turned out to be correct.
Readers desirous of learning more about the recent developments in
solar dynamo and about the flux transport dynamo model may turn to
the review articles by \citet{Charbonneau10,Charbonneau14} and \citet{Chou11,Chou13,Chou23}.

The $\alpha \Omega$ dynamo model proposed by \citet{Parker55a} was
undoubtedly the most important single step in our theoretical
understanding of the solar dynamo, even though we now believe 
that the original model has to be modified in significant ways. 
Although the $\alpha$-effect may be suppressed in the regions
inside the Sun where a strong toroidal magnetic field is produced
by differential rotation, there is no doubt that this is a real
effect operative in many astrophysical systems and has been supported
by numerical simulations.  There have been many impressive simulations of
the geodynamo starting from the work of \citet{Glatz95}.  These simulations
show the $\alpha$-effect at work---exactly the way Parker envisaged it
many decades ago.

\section{MHD of the outer solar atmosphere}\label{sec5}

Parker wrote more papers on the solar corona (if we include
his solar wind papers also in this category) than on any
other astrophysical topic throughout his career. Certainly
the solar corona provides illustrations of many processes
important in plasma astrophysics. Since the emission from
the corona is much feebler than the emission from the solar
surface, the corona is not visible from the Earth's surface
under normal circumstances and has been studied historically during total
solar eclipses. In the 1930s, \citet{Lyot39} developed the
coronagraph which produces an artificial eclipse 
inside the telescope by blocking light from the
solar disk. However, diffuse light still makes it very
difficult to see the corona and a coronagraph has to be
taken to a high mountain to get a glimpse of the corona.
From the 1970s, it has been possible
to observe the solar corona continuously with the help of coronagraphs
carried in space missions. Because of the high
temperature, the corona emits copious amounts of X-rays.
X-ray imaging instruments sent to space (starting from
Skylab in the early 1970s) have provided increasingly striking
images of the X-ray emitting regions of the corona.  X-ray
emissions detected from many solar-like stars indicate that
they also have similar stellar coronae.

Even a visual inspection of the structures in the solar
corona suggests that magnetic fields must be behind them.
After the development of the idea of magnetic buoyancy
\citep{Parker55a}, it became clear that the magnetic fields
created within the solar convection zone would come out in
the corona.  Although a direct measurement of the magnetic
fields in the corona has proved particularly challenging,
solar astronomers started gathering different kinds of
evidence within the last few decades that the corona is full
of magnetic fields.  We have already pointed out in the beginning
of section~4 that the plasma-$\beta$
is expected to be less than 1 in many regions of the lower
solar corona, indicating that magnetic fields would control
the dynamics in those regions.

Parker's important contributions to coronal physics
can be broadly classified under three topics: (i) the basic
theory of magnetic reconnection; (ii) the theory of coronal
heating and (iii) the theory of solar wind.  If we wanted to
present our discussion of these topics in a chronological order
depending on the time when Parker worked on these topics, 
then we have to put topic (ii) after topic (iii), since
Parker worked extensively on some of the theoretical aspects
of the coronal heating problem in the last few years of his
active scientific career.  However, we have decided to follow
an order which appears more logical to us.  After all, it is
the hot corona which drives the solar wind.

\subsection{Solar flares and magnetic reconnection}

The importance of magnetic fields in the dynamics of the solar corona first
became apparent from the study of solar flares, which are
gigantic explosions taking place above the solar surface.
A powerful flare may release energy of order $10^{32}$ erg
within a time of the order of an hour. The crucial issue was
to identify this source of energy.  The first recorded flare
observed by \citet{Carring1859} occurred over a sunspot. Further
observations over the next few decades established that
flares take place above solar active regions, which are
expected to be dominated by magnetic fields ever since the
discovery of magnetic fields in sunspots \citep{Hale1909}.
It was natural to guess that the magnetic energy would be the
source of the energy released during a solar flare.  The
important question was to work out the detailed physics of
the mechanism by which this energy conversion takes place.

It is clear from (2) that the term $\eta \nabla^2 \Bb$ corresponds to the magnetic field (i.e.\ the magnetic energy)
getting dissipated due to the resistivity.  In 
coronal plasma with very low resistivity, this
term can be significant only if $\nabla^2 \Bb$ is large.
After \citet{Dungey53} pointed out the importance of magnetic neutral
points, \citet{Sweet58} realized that  $\nabla^2 \Bb$ would be large
if we have a null surface with oppositely directed magnetic
fields on the two sides. Since Amp\'ere's law
suggests that the current density would be very
high at the null surface, such surfaces are
often referred to as {\em current sheets}. What is more, if the magnetic
field at the current sheet is dissipated without any motions
within the plasma, then that would cause a decrease in
the magnetic pressure, leading to a pressure imbalance.
This suggests that plasma with oppositely directed
magnetic fields on the two sides of the neutral surface (or the current sheet)
would flow into the region of the neutral surface.  As a result
of this, the process can continue as long as we have
fresh magnetic fields brought from the two sides by the inflowing
plasma. Based on Sweet's ideas, \citet{Parker57} managed to estimate
the velocity with which the inflowing plasma would move
towards the current sheet.  This inflowing velocity
is known as the {\em Sweet-Parker reconnenction rate}.
Since the plasma which is squeezed out of the region of the
neutral surface contains magnetic field lines, parts of which 
were originally on the two opposite
sides of the neutral sheet, this process came to be known
as {\em magnetic reconnection}. It may be worthwhile to point out a little bit of the interesting publication history. Sweet presented his work at the International
Astronomical Union Symposium 6 held in Stockholm during
27--31 August, 1956. Sweet's paper came out in the 
Proceedings of Symposium, which was eventually published
only in 1958 \citep{Sweet58}.  Due to this delay in the publication
of Sweet's paper, Parker's paper appeared in print
earlier \citep{Parker57}. However, \citet{Parker57} generously titled the paper
``Sweet's mechanism for merging magnetic fields in 
conducting fluids'' and pointed out in the opening sentence
that he was elaborating on Sweet's ideas.

Soon after the idea of magnetic reconnection was put
forth, it was realized that this is an extremely important
process in many plasma systems in the laboratory and in
many astrophysical systems.  However, the Sweet-Parker
reconnection rate seemed inadequate to explain the rather
short rise-phase and duration of solar flares.  The reconnection has
to proceed at a much faster rate for release of such
a substantial amount of energy in such a short time
during a typical solar flare.  With this realization,
the search for reconnection at a faster rate began.
One of the first influential scenarios for faster
reconnections was proposed by \citet{Pets64}.
As numerical simulations of reconnection started,
it was apparent that the reconnection rate may depend on
boundary conditions far away---making this a particularly
challenging problem, which depended not only on the local
conditions, but also on what was happening far away. Readers
desirous of learning more about this complex subject are
referred to the monographs by \citet{Priest00} and \citet{Priest14}
and the living review article by \citet{Pontin22}. 
For a discussion of the physics behind solar
flares, see \citet{Shibata11} and \citet{Priest14}. We shall
now turn our attention to another subject in which
magnetic reconnection plays a crucial role.

\subsection{Coronal heating}

While the temperature of the solar surface is
about 5800 K, the temperature in certain regions
of the corona can be as high as (1--2) $\times 10^6$ K.
It was first realized in the 1940s that the corona
is much hotter than the solar surface.  Some emission
lines seen in the solar corona were identified by
\citet{Edlen43} as lines produced by iron
atoms which have lost several electrons.  Such a
loss of several electrons from iron atoms would
be possible only if the corona had a very high
temperature.  What produces the high temperature
of the corona became a central question of theoretical
solar  physics.  \citet{Bier48} and \citet{Schwarz48} were the
first to suggest that acoustic waves produced
by convective motions just below the solar surface
could propagate to the corona and dissipate there to
produce the high temperature, whereas \citet{Alf47}
proposed that MHD waves do this job.  As the magnetic nature
of the corona became more apparent, it was realized
that we need to consider MHD waves propagating in
the corona rather than simple acoustic waves.  In
order to produce the high temperature of the corona,
the MHD waves have to dissipate in the corona
rather than passing through it without much dissipation,
which would  be the case if the resistivity of the corona
was too low (as expected).  Mechanisms such as
{\em phase mixing} and {\em resonant absorption} have been
suggested to enhance the dissipation.  We shall
not enter into a detailed discussion of this vast
subject. Interested readers are referred to a brief
discussion in \citet{Priest14}, pp.\ 356--364.

\begin{figure}
    \centering
\includegraphics[width= 0.7\textwidth]{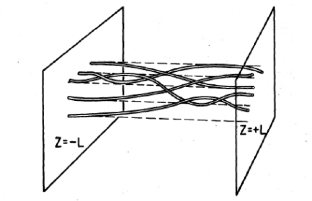}
    \caption{A famous figure from the coronal heating paper \citep{Parker72} showing how magnetic field lines between two parallel planes develop a complex topology due to the motions of footpoints. Reproduced by permission of the AAS.}
\end{figure}

\citet{Parker72} suggested an alternative mechanism
of how the heat is produced in the corona.  It
is expected that magnetic buoyancy would make the
magnetic field rise through the convection zone to
come out in the corona in the form of magnetic
loops.  When X-ray images of the corona from
space missions first
became available in the 1970s, it was indeed found
that the corona is full of X-ray emitting loops.
The enhanced emission from these loops suggested
that the magnetic loops are the hottest regions
of the corona and that these loops are the
primary locations in the corona within which the heat
is produced.  One may naively expect that the magnetic
fields inside coronal magnetic loops would be in
magnetostatic equilibrium.  However, these loops
are not isolated systems.  Magnetic field lines
in these loops must continue below the photospheric
surface of the Sun.  The convective motions present
there are expected to move the magnetic footpoints 
of the loops, thereby disturbing the equilibrium of the
the overlying magnetic structures. As a result, the
magnetic fields in the loops would try to relax to new
configurations satisfying magnetostatic equilibrium.
On the basis of a very short mathematical derivation,
\citet{Parker72} argued that the magnetic fields
would relax to configurations having discontinuities
in the magnetic field, i.e.\ would give rise to
current sheets where magnetic reconnection can
take place to produce the heat.

Parker often told many of us that the 1972 paper
on coronal heating was his own favourite among
all his papers.  This certainly turned out to be
his most controversial paper.  Many peers in the
community were not sure whether the argument \citet{Parker72}
gave on the basis of a short derivation was sufficiently
sound.  It is worthwhile to reproduce the short
derivation here. 
\citet{Parker72} realized that the curvature
of the coronal loop was not essential in this problem.
So he considered a uniform initial magnetic field in one direction
(say the $z$ direction) between two perpendicular
plane surfaces (representing the two ends of the
loop at the photospheric surface). Now suppose that
random motions on the plane surfaces perturb
the magnetic field, as sketched in Figure~8.  
The perturbed magnetic field
would try to relax to a configuration satisfying
the magnetostatic equilibrium equation, which we arrive
at by putting $\vb = 0$ in (1) and is
$$- \frac{1}{\rho} \nabla \! \left(p+ \frac{B^2}{8 \pi}\right) 
+ \frac{(\Bb. \nabla)\Bb}{4 \pi \rho} = 0. \eqno(6)$$
We now write the magnetic field as
$$\Bb = B_0 {\bf e}_z  + \bb, \eqno(7)$$
where $B_0 {\bf e}_z$ is the initial magnetic
field assumed uniform and $\bb$ is the perturbation produced in
it by footpoint motions. On substituting (7) in (6) and
keeping only terms linear in $\bb$, we obtain
$$- \frac{1}{\rho} \nabla \! \left(p+ \frac{B_0 b_z}{4 \pi}\right) 
+ \frac{B_0}{4 \pi \rho} \frac{\pa \bb}{\pa z} = 0. \eqno(8)$$
On taking the divergence of this equation and keeping
in mind that $\nabla. \bb = 0$, we arrive at
$$ \nabla^2 \left(p+ \frac{B_0 b_z}{4 \pi}\right) = 0,
\eqno(9)$$
which is the Laplace equation, one of the most thoroughly
studied equations of mathematical physics.  If we demand
that the solution of this equation must not blow up
anywhere (including infinity), then the only possibility
is that the solution is spatially constant. We expect
(9) to hold between the two parallel plates where we
rule out the possibility of the solution blowing up anywhere.  This
forces us to the conclusion
$$p+ \frac{B_0 b_z}{4 \pi} = {\rm constant}.$$
It then follows from (8) that
$$\frac{\pa \bb}{\pa z} = 0. \eqno(10)$$
This implies that magnetostatic equilibrium requires an
invariance along a symmetry direction.

We expect everybody to agree with this derivation 
so far.  The uncertainly arises when we try to
interpret (10). If we consider the idealized case
of a plasma with zero resistivity, then the only
term on the right hand side of (2) is 
$\nabla \times (\vb \times \Bb)$, which would
imply that the magnetic field would move with the plasma
and magnetic topologies cannot change. 
In general,
we expect the topology of the magnetic field lines
resulting from arbitrary footpoint motions to be such
that it may not be possible to satisfy (10).  In other
words, we seem to have two requirements which are difficult
to reconcile.  On the one hand, the magnetostic equilibrium
demands that (10) be satisfied.  On the other hand, topological
constraints may make it impossible to satisfy (10).  What then
happens? \citet{Parker72} argued that the magnetic field
would relax to a configuration with internal discontinuities
where the magnetic field would cease to be
continuous and differentiable.  In other words,
many current sheets may arise. Magnetic reconnection
would take place at these current sheets and heat is expected
to be generated to cause the high temperature of the
corona.

Do we find this argument convincing?  It is no wonder that
many in the scientific community felt somewhat unsure
about this argument.  A review article on coronal heating
mechanisms published in 1981 \citep{Kuperus81} did not even
cite Parker's 1972 paper! A great deal of interest
was again re-kindled in this subject from the early
1980s when the space-based Einstein X-ray Observatory
established that many solar-like stars have coronae
emitting X-rays \citep{Pall81}. Based on
some reasonable assumptions, \citet{Parker83} estimated
the amount of energy expected to be generated due to the
formation of magnetic discontinuities in the solar corona
and found that it approximately matches the energy budget needed
to heat the corona.  \citet{Parker88} also argued that the magnetic
discontinuities would give rise to many small reconnection
regions, which he called {\em nanoflares}, rather than one
large reconnection region as in a large flare.  Since the
dissipation of MHD waves in the corona had been proposed
as another mechanism for coronal heating, a debate took
place in the 1980s and 1990s as to which of the two 
mechanisms---dissipation of MHD waves and current sheet
formation due to footpoint motions---was the correct
mechanism for coronal heating.  A view has emerged gradually
over the years that both these mechanism must be at work
in different regions of the corona.  Current sheet
formation is possible only in regions of closed magnetic
field, such as coronal loops.  They are also the hottest
regions of the corona, suggesting that a different heating
mechanism may be at work inside them.  Presumably the coronal
loops are heated by the formation of many small current sheets
as envisaged by Parker, whereas the other regions of the corona
with open magnetic field lines are probably heated by the dissipation
of MHD waves.

Apart from applying his theory to several aspects of observational
data, Parker was concerned with the question whether
one could justify the arguments of his 1972 paper by
more rigorous detailed calculations.  A heated debate
took place on this subject in the mid-1980s within the
American solar physics community when \citet{vabB86}
claimed that the invariance condition (10) may not be
essential for magnetostatic equilibrium. However, his analysis
pointed out that the braiding of field lines by footpoint
motions may lead to a cascade of energy to smaller
length scales, as envisaged by Parker. Parker himself
wrote a series of papers in the late 1980s and the early
1990s advancing various kinds of arguments to justify
the suggestions he made in his 1972 paper. For example,
he pointed out that the mathematical theory of the structure
of field lines would be analogous to the mathematical
theory of light rays in a medium of varying refractive
index \citep{Parker89}.  Parker exploited this analogy to draw various
conclusions. 

Perhaps Parker had not struggled with any other
scientific question as much as he struggled 
in the later part of  his scientific career to
understand the mathematical structure of the theory
of magnetostatic equilibria and to address the
question whether the theory inevitably leads to
the conclusion that magnetic discontinuities must
arise in the general situation.  It is a difficult
subject and it has to be admitted that Parker did
not attract many followers.  In other words, there
were not many in the scientific community to take
the lead from Parker's work to carry on further investigation
of this subject.  As a result, the majority of Parker's
many papers dealing with these basic theoretical issues 
written after mid-1980s
received relatively few citations, although some of
the papers in which he discussed the application
of his theoretical ideas to explain the observations
of solar and stellar coronae became citation
classics \citep{Parker88}!  When Parker eventually felt
that he had succeeded in developing a unique perspective
of the subject, he decided to put forth a coherent 
account of the subject in his monograph {\em Spontaneous
Current Sheets in Magnetic Fields} \citep{Parker94}.  It is certainly
not an easy book to read---in contrast to Parker's
earlier monograph {\em Cosmical Magnetic Fields}
\citep{Parker79a},
known for its remarkable lucidity, which is a pleasure
to read because of its elegant writing style.  Arguably,
Parker had to deal with an intrinsically difficult 
subject in his later monograph. We do know of a few 
examples of scientific investigation which did not get
much attention from the contemporaries, but led to
important developments many years after the work.  Still,
when the sustained efforts of a great scientist over
many years do not get too much attention from 
contemporaries, one cannot avoid asking an awkward
question.  Were the scientific returns commensurate
with the time and energy Parker spent on this subject?
I would  humbly submit that my own understanding of
this complicated subject is very limited and I am
not qualified to answer this question. I refer
the readers to a review by \citet{Low23}, who 
delivered the prestigious Crafoord Prize lecture
on behalf of the ailing Parker (who won this Prize
in 2020) and had many discussions on the 
magnetostatic theorem with Parker in his declining
years.

\subsection{Solar wind}

We now come to Parker's most famous work: the prediction
of the solar wind.  At a time when the solar corona was known
to be very hot but there was not much understanding about the
reason behind this, \citet{Parker58} pointed out that a hot corona
would drive an outward flow of plasma through the solar system.
There has never been a more radical transformation in our view
of the space environment of our planet Earth. The prevalent
view for several centuries was that the interplanetary 
space is essentially empty, through which the planets
encircle the Sun. Parker's work suggested that the Earth
is basically embedded in the extended atmosphere of the Sun,
leading to the possibility of understanding how phenomena
on the Sun (like solar flares) may affect the Earth.

To explain why comet tails turn away from the Sun,
\citet{Bier51} suggested that there may be a corpuscular outflow
from the Sun which turns the comet tails in the outward
direction with respect to the Sun. On the other hand,
\citet{Chap57} pointed out that the high temperature of the
corona suggested that the corona would extend to a very
large distance from the Sun.  Now, it is not possible for
a stream of plasma to flow through a background of plasma at rest.
That would lead to two-stream instability.  Parker realized
that the outer parts of Chapman's extended corona must expand
to produce Biermann's corpuscular outflow.  From the
basic equations, \citet{Parker58} was able to find a solution
which exactly corresponded to this situation.  Parker's
original calculation was essentially hydrodynamic, although
in the later part of the paper he discussed how the wind would
affect the magnetic field coming out of the Sun.  In fact,
the calculations are so straightforward that a perusal
of Parker's paper may give the misleading impression that
this work could be done by an average scientist of much lesser
abilities.  However, only Parker had the great insight to
look at this problem in this particular way.

\begin{figure}
    \centering
\includegraphics[width= 0.6\textwidth]{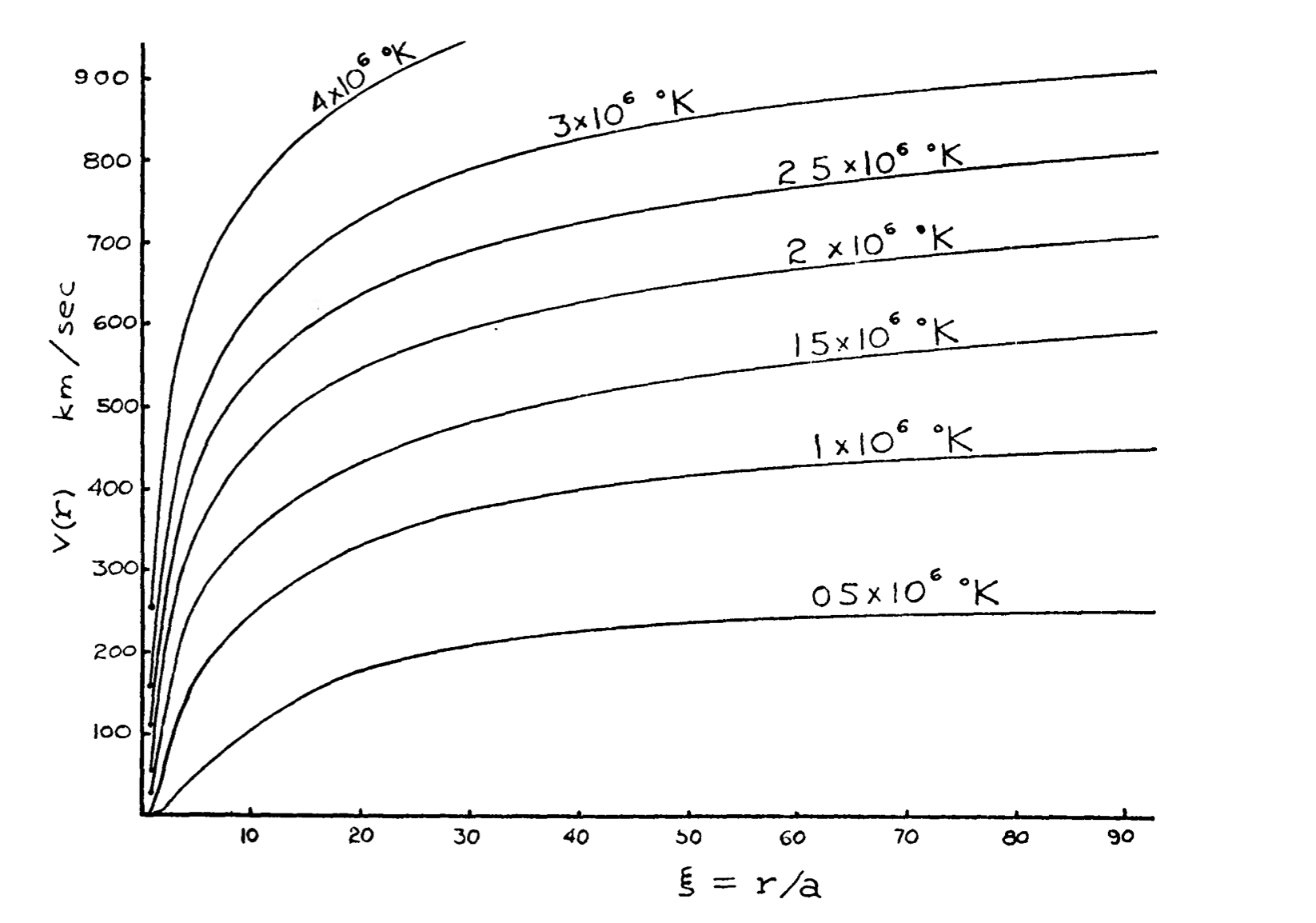}
    \caption{A figure from the solar wind paper \citep{Parker58}  showing  how the speed
    of the solar wind was found to vary with radial
    distance from the Sun on the basis of the theoretical model for different assumed values of the temperature. Reproduced by permission of the AAS.}
\end{figure}

Assuming spherical symmetry, if 
the radially outward velocity is $v$ at a radial distance $r$
where the density is $\rho$, mass conservation suggests that
$$ \rho v r^2 = {\rm constant}. \eqno(11)$$
To consider a static hydrodynamic flow, we need to use 
the radial component of (1)
by putting the time derivative term and the magnetic force
terms to zero.  This gives
$$\rho v \frac{d v}{d r} = - \frac{dp}{dr} - \frac{G M_{\odot}}{r^2}
\rho, \eqno(12)$$
where we have put the appropriate expression for the Sun's gravitational
field in the force term. It is clear that the two scalar
equations (11) and (12) involve three independent 
variables---namely $\rho$, $p$ and $v$---
which would all be functions of $r$ alone in a spherically
symmetric situation.
If one can relate $\rho$ and $p$, then the number of independent
variables would be equal to number of equations, which can then
be easily solved.  \citet{Parker58} made the simplest assumption
of an isothermal condition.  On solving (11) and (12), he was
able to find solutions involving flows which start from very
low subsonic velocities 
near the solar surface and eventually become supersonic at
some distance away from the Sun. Figure~9 shows some solutions obtained
by Parker for the variation of velocity $v$ with the radial
distance $r$ for different assumed values of
the temperature of the corona. 
While the solar wind is expected to stretch out any
magnetic field lines coming out of the Sun, \citet{Parker58}
realized that the solar rotation would impart a spiral structure
to the field lines in the equatorial plane of the Sun,
as shown in Figure~10.  These
spirals are now referred to as {\em Parker spirals}.  Whether
a solar explosion would affect the Earth often crucially depends
on whether the site of the explosion and the Earth lie close to one
Parker spiral. \citet{Parker58} estimated that the mass loss of the
Sun during its lifetime due to the solar wind would be negligible.
However, the magnetic field stretching out of the Sun would be an
efficient transporter of angular momentum and the Sun might have
lost a significant amount of angular momentum taken away by the
solar wind, implying that the Sun might be rotating faster when
it was young. The basics of solar wind theory are discussed
so beautifully and coherently in Parker's original paper that it
can still be recommended as a pedagogical introduction to students who
want to learn the subject!

\begin{figure}
    \centering
\includegraphics[width= 0.6\textwidth]{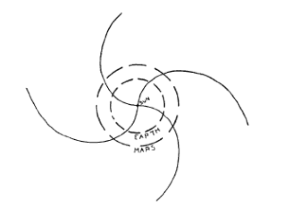}
    \caption{A sketch from the solar wind paper paper \citep{Parker58} showing how the magnetic
    field lines of the rotating Sun would get stretched due to the
    outflowing solar wind, giving rise to what are
    now called Parker spirals. Reproduced by permission of the AAS.}
\end{figure}

Parker became interested in this subject during a visit of Biermann
to Chicago and then had a discussion with Chapman when he visited
the High Altitude Observatory in Boulder where Chapman was working.
Parker has given an account of the dramatic history behind the discovery
of the solar wind in a {\em Scientific American} article \citep{Parker64}.
It is well known that the solar wind paper was rejected by two
referees. Here is Parker's own account of the publication history
\citep{Parker2014} (SCR refers to solar corpuscular radiation):

\begin{quote}
I wrote it up for publication in
The Astrophysical Journal, of which, fortunately, Prof. Chandrasekhar was editor at that time. The
referee’s report came back in a few months with the suggestion that the author should spend some
time in the library to familiarize himself with the SCR before attempting to write a scientific paper
on the subject. There was no specific criticism of the mathematics or of the interpretation of the
observations. So Chandra sent the paper to a second “eminent” referee, with essentially the same
result. I emphasized to Chandra that these two referees, for all their hostility, could find no scientific
error. Then one day Chandra came to my office and said, “Now see here, Parker, do you really want
to publish this paper? I have sent it to two eminent referees, and they both say the paper is wrong.” I
replied that the referees had no scientific criticism. He thought for a moment and then said, “Alright,
I will publish it.” Some years later he told me that he had been skeptical about the paper, but without
objective criticism, he felt obliged to publish it. To my regret I failed to save the two referee reports \ldots
\end{quote}
Chandrasekhar agreed to publish the paper in spite of his own reservations.
However, Joseph Chamberlain, another colleague of Parker working at the
Yerkes Observatory belonging to the University of Chicago, was convinced
that Parker's theory could not be correct and wrote a paper pointing out
why he considered it wrong \citep{Chamber60}. Figure~11
shows a part of Parker's letter commenting on the solar
wind soon after he had developed its theory. Parker's generous
nature is rather evident in this letter.  Instead of bragging
about his own work, he gives a lot of credit to Biermann.

The theory of the solar wind was generally accepted only after in-situ
measurements from space vehicles confirmed its existence.  It is a
remarkable historical coincidence that Parker's theory of the solar wind
was worked out almost exactly at the time when the first artificial
satellite Sputnik was launched (on 4 October 1957), heralding a space
race between the USSR and USA. The first detection of the solar wind
was made by the Russian spacecraft Luna-2 \citep{Grin60}. It may
be kept in mind that this was the era of the infamous cold war which reached its peak
during the Cuban missile crisis of 1962. It is quite remarkable that Russian
space missions were busy confirming the theory of an American scientist
during the height of the cold war.  What better example can one give of
international co-operation in science! An interesting Russian perspective
of the history of the solar wind can be found in \citet{Obridko17}.

\begin{figure}
    \centering
\includegraphics[width= 0.95\textwidth]{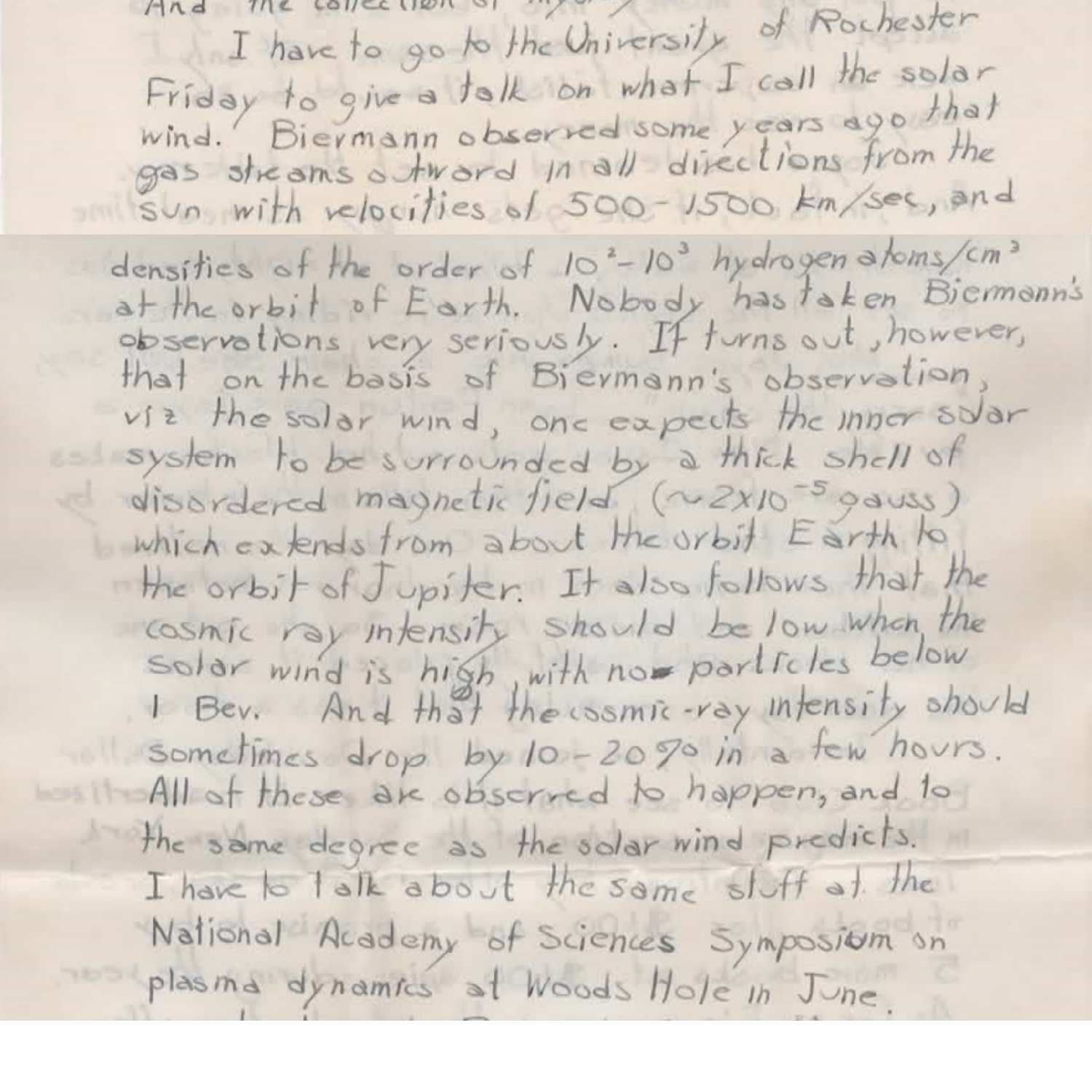}
    \caption{A part of Parker's letter to his parents
    describing the solar wind. This letter is dated
    16 February 1958, which means that it was written
    only a few weeks after the famous paper on solar
    wind was submitted to {\em The Astrophysical Journal} (on 2 January 1958).
    Credit: Eric Parker and Susan Kane-Parker.}
\end{figure}

After the existence of the solar wind was established, \citet{Parker63a}
wrote a monograph on the subject.  He also pointed out the possibility of
similar winds in other stars and discussed the extension of the theory
when one goes beyond the isothermal assumption in a comprehensive review
of the subject \citep{Parker1965}.  After this review presenting his final point of view,
he almost left the subject of solar wind theory and did not work on this subject much
after that.  The only aspect of the solar wind on which he worked extensively beyond the
mid-1960s was the propagation of cosmic rays through the solar wind
to be discussed in subsection~6.1. This contrasts strikingly with Parker's
engagement with the coronal heating problem till the end of his active
research career.  Perhaps he felt that some of the fundamental questions
connected with the coronal heating problem had not been answered
satisfactorily and he wanted to develop a deeper understanding of
the subject.  On the other hand, while many important research questions
kept coming up in the field of solar wind, probably Parker felt that
there were no such major unsettled conceptual issues in that field
and he left the field which he single-handedly established for others
to take forward.

I shall not try to present a comprehensive account of the later developments
in solar wind theory, for which I refer the reader to
\citet{Priest14}, Chap. 13. Here I shall only make a few comments on
the works which extended Parker's original work.  \citet{Parker58}
considered the effect of the solar wind on solar magnetic fields,
but did not develop a full MHD theory combining the gas and the
magnetic field.  This was done by \citet{WD67} for the equatorial
plane and was extended by \citet{Saku85} beyond this plane.  
One consequence of an MHD wind pointed out by \citet{Parker58}
is the magnetic
braking of rotation, as we have mentioned.  The importance of
such braking of rotation for different kinds of stars was recognized
soon after Parker's work on the solar wind \citep{Schatz62,Mestel68}. 
It has been
realized from total solar eclipse photographs that closed magnetic
regions in the corona often give rise to helmet-like structures, with
the solar wind flowing by their sides.  Such structures were modelled
by \citet{Pneu71}. As more and more X-ray images of the corona started
coming from space missions, it became clear that the corona is anything
but spherical, pointing out the need to go beyond the spherically
symmetric model of the solar wind developed by Parker.  X-ray images
indicated the existence of dark coronal holes and the solar wind
emanating from such holes was found to be more energetic.  It became
clear that some energy must be getting deposited in coronal holes
in the right manner to produce a more efficient acceleration of the
solar wind in those regions.  This subject has been reviewed by
\citet{Leer82}. Let us end this discussion by pointing out that the
Parker Solar Probe, named in honour of Parker and launched while
he was still alive, is now exploring the regions of the corona from
which the solar wind emanates. See \citet{Raou23} for a discussion
of the science results obtained by this mission.

\section{Other significant solar physics works}\label{sec6}

After summarizing some of Parker's most famous works on solar
physics, I shall now briefly discuss some of his other important
works in the field.

\subsection{Theory of cosmic ray propagation}

It was in the 1960s that Parker made fundamental contributions
in the theory of the propagation of cosmic rays through the solar
wind.  By that time, the existence of the solar wind had been
firmly established and it was realized that cosmic ray
particles have to make their way through the solar wind to 
reach the Earth.  All the other solar physics works of Parker
which we summarize were based on the macroscopic MHD equations
(1) and (2)---except his work on cosmic rays, for which he
followed a more microscopic approach as we shall discuss now.
Since my own knowledge of this subject is very limited, I
shall restrict myself only to a few broad remarks and
refer the interested reader to \citet{Jokipii71} for a rigorous review
of how the field of cosmic ray propagation developed in the
1960s.

Cosmic rays were discovered by \citet{Hess12}.  By the middle of the
twentieth century, there were enough indications that the
cosmic ray flux reaching the Earth was affected by solar activity.
\citet{Forbush54} discovered that there is a dip in the cosmic ray flux
after a major solar flare.  With more data of the cosmic ray
flux gathered over the years, there was also the indication of
an anti-correlation with the solar cycle---the cosmic ray flux
decreasing at the time of the sunspot maximum.  It was clear
that enhanced magnetic activity within the solar system made
it more difficult for cosmic ray particles to reach the Earth.
The important scientific question was to provide a proper
theoretical framework to understand this.

To explain how the charged particles making up cosmic rays
get accelerated to very high energies, \citet{Fermi49} proposed a famous
mechanism involving interstellar gas clouds 
with magnetic fields which act as magnetic
mirrors and reflect the gyrocentres of moving charged particles.
According to the original theory of \citet{Fermi49}, the charged particles
are accelerated by repeated reflections from randomly moving 
interstellar gas clouds. While Fermi's idea of charged particles
getting reflected from magnetic irregularities turned out to be
very influential, it was realized by the late 1970s that blast
waves emanating from supernova explosions provide more efficient
sites of particle acceleration than moving interstellar gas clouds.
Parker carried out his research on cosmic rays at a time when
supernova explosions had not yet been identified as the sources of
cosmic rays.  However, there was already a widely held view that
the majority of 
cosmic ray particles come to the solar system from interstellar
space.

After the discovery of the solar wind, \citet{Parker65} realized that the
outflowing solar wind would carry turbulent magnetic fields with
it and the irregularities in these magnetic fields would act as
scattering centres for moving charged particles. The cosmic ray
particles have to diffuse through the magnetic irregularities of
the solar wind, while being advected with the velocity of the solar
wind because of the advection of the magnetic scattering centres
with the solar wind.  \citet{Parker65} showed that the time evolution of the density of 
the cosmic ray particles would be governed by the Fokker--Planck
equation. To make quantitative calculations, the crucial quantity
one had to estimate was the diffusion coefficient arising out of the
repeated scatterings of the charged particles by the magnetic irregularities
in the solar wind. Since the charged particles are expected to
diffuse more easily parallel to the large-scale magnetic field
of the solar wind than perpendicular to it, the diffusion
coefficient is expected to be anisotropic---the coefficient for
diffusion parallel to the magnetic field being larger 
than the coefficient
for diffusion in the perpendicular directions.  From reasonable
assumptions about magnetic irregularities in the solar wind, \citet{Parker65}
estimated the diffusion coefficients and found them to be in broad
agreement with experimental data of cosmic rays.

Later, \citet{Parker69} developed a more complete theory of
how the turbulent velocities in the solar wind would produce
stochastic fluctuations in the magnetic field.  They realized
that the charged particles would tend to gyrate around the 
large-scale magnetic fields of the Parker spirals.  However,
due to the scattering from magnetic irregularities, there would
be continuous spreading of the cosmic rays in the perpendicular
direction.  Their calculations of the rate of this perpendicular
spreading agreed with experimental data. 

\subsection{Magnetic flux tubes in the solar convection zone}

It is one of the remarkable observational facts that the
magnetic fields at the solar surface appear concentrated
within structures of different sizes.  Sunspots are the
largest concentrations of magnetic flux. Big sunspots with
magnetic field of about 3000 G can sometimes be so large that
it may be possible for the whole Earth to be immersed in
one of them. Solar astronomers discovered that magnetic
fields outside sunspots exist in the form of smaller 
magnetic flux tubes having sizes of a few hundred km with
magnetic field of the order of 1000 G inside them \citep{Sten73}.  
Understanding why magnetic fields at the solar surface
exist in the form of flux tubes of different sizes has
been a challenge for theoretical MHD and a topic which
interested Parker greatly.

\citet{Bier41} explained the darkness of sunspots by suggesting
that the tension of the magnetic field---which arises
out of the term involving $(\Bb.\nabla)\Bb$ in (1)---inhibits
convective heat transport inside sunspots so that sunspots become
cooler than the surroundings.  The linear theory of 
{\em magnetoconvection} (i.e.\ convection in the presence of
magnetic fields) was developed by \citet{Chandra52},
showing that the magnetic field indeed inhibits convection.
If magnetic fields are concentrated in some regions inside
a convective gas, we certainly expect the convection to
be inhibited within those regions.  But why should the
magnetic field be concentrated in some regions? To
address this  question, \citet{Parker63b} carried out some
elegant mathematical calculations to study how magnetic
fields evolve in regions of stationary fluid flows
having some circulatory patterns inside them.  He found
that magnetic fields tend to get swept away from 
interior regions of circulatory fluid flow and are concentrated in regions of converging fluid flow, which
\citet{Parker63b} had taken to be vertical.  Within a few years
of this, \citet{Weiss66} carried out a numerical simulation to
confirm Parker's idea.  These demonstrations that magnetic
fields become concentrated within regions of converging flow
provided the important first step towards understanding
how concentrations of vertical magnetic field arise at the
top of the solar convection zone.

If magnetic fields are concentrated by converging flows,
it is easy to argue that the magnetic energy density $B^2/8 \pi$
should at most be of order of the fluid kinetic energy
$(1/2) \rho v^2$. However, observational studies of small magnetic flux tubes at the solar surface indicated that
the magnetic fields inside them have energy
densities a few times
larger than the kinetic energy density of surroundings
fluids.  It was clear that some additional mechanism was
needed to concentrate the magnetic field further. \citet{Parker79b}
showed how this may happen by considering a vertical
magnetic flux tube in hydrostatic equilibrium. \citet{Parker79b}
argued that downward fluid flows inside such flux tubes
may give rise to an instability, leading ultimately to
a different configuration of the flux tube with stronger
magnetic field inside.  This process has been named 
{\em convective collapse}. Further analysis of this subject
was presented by \citet{Spruit79} and a numerical simulation was
carried out by \citet{Hasan85}.

\begin{figure}
    \centering
\includegraphics[width= 0.6\textwidth]{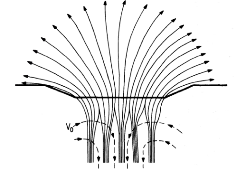}
    \caption{The fibril structure of sunspots proposed
    by \citet{Parker78}. Reproduced by permission of the AAS.}
\end{figure}

One other influential work of Parker connected with flux
tubes which we would like to mention is his theory of the
structure of sunspots.  \citet{Parker78} pointed out that many aspects
of observational data about sunspots can be explained better
if a sunspot is a collection of magnetic flux tubes held
together rather than a monolithic flux tube.  Figure~12 sketches
the structure of sunspots \citet{Parker78} proposed. Further support for
this model came when observational study of the emergence of
sunspots showed that sunspots often form through the process
of smaller flux tubes emerging first and then coming 
together \citep{Zwaan85}.

\section{ Parker's important non-solar contributions}\label{sec7}

As already pointed out, Parker had broad interests in plasma
astrophysics and did not want to be identified merely as a solar
physicist.  We have repeatedly emphasized that even Parker's works
described in sections~4--6 can be readily applied to those solar-like stars 
which have magnetic activity like the Sun. 
We now turn our attention to Parker's other important
contributions beyond solar physics. 

Before discussing specific research contributions, we take note
of the monumental 800-page monograph {\em Cosmical Magnetic
Fields} \citep{Parker79a}.  Although this large book packed with equations may
appear forbidding at first sight, it provides pleasurable
reading because of its elegant and clear style of writing.
A large part of this classic of plasma astrophysics is devoted
to developing many of the basic topics of MHD which have the
possibility of wide applications to various astrophysical systems.
After developing the basics, \citet{Parker79a} considers applications to
planets, stars and galaxies in the last few chapters.  It should
be pointed out that to some extent the choice of topics was guided
by Parker's own research interest and not all types of astrophysical
magnetic fields are covered in this book.  For example, there is
no discussion about pulsars and their magnetic fields. 
Although Parker himself had analyzed the problem of magnetic braking
of the rotating Sun by the solar wind and it was recognized that magnetic braking of protostars is extremely important in the star
formation process, curiously there is no discussion of magnetic
braking in Parker's book.  Some of the important topics of
stellar magnetism not included in Parker's book have been discussed in
the book by \citet{Mestel99}, another classic volume in the same
{\em International Series of Monographs on Physics} in which Parker's
book had appeared.
It may be
mentioned that Parker's book was translated into Russian in two 
volumes---Volume I translated by A.\ Ruzmaikin and Volume II by
A.\ Shukurov. The 2-volume set was edited by Ya B.\ Zeldovich
and published in 1982 by Mir Publishers (information about this translation
was provided to me by Anvar Shukurov).  Parker received 
these volumes when I was a student in the group.  Although
he normally would not display his emotions, he was as happy
as a child to receive these two volumes and excitedly
brought them to my office to show me. 

\begin{figure}
    \centering
\includegraphics[width= 0.5\textwidth]{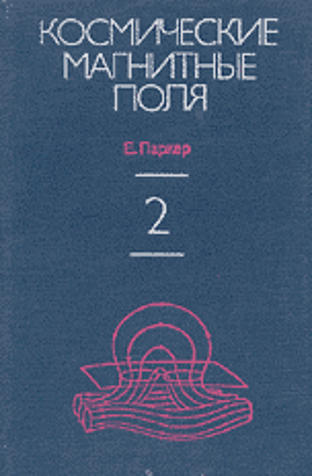}
    \caption{The cover of Volume 2 of the Russian
    translation of Parker's book {\em Cosmical Magnetic
    Fields}.}
\end{figure}
Since much of Parker's non-solar work deals with the magnetic
field in the interstellar medium of galaxies, let us first say a few
words about the initial history of the subject.  Parker himself has
provided an account of this early history from his perspective: \citet{Parker79a}, pp. 795--807.  The first indication that
a magnetic field spans our Galaxy came when \citet{Hilt49}
discovered the light from many stars to be polarized.  \citet{Davis51}
pointed out that a galactic magnetic field must
have aligned paramagnetic dust grains so that the interstellar
medium acts as a polarizer. \citet{Chandra53}
were to present one of the first estimates
of the strength of the galactic magnetic field based on
ingenious theoretical arguments: of order
$10^{-6}$ G. From the polarization data of many stars, it
could be inferred that the magnetic field is in the
direction of the spiral arm of the Galaxy. One interesting
fact became apparent even in the early days of research.
The energy densities of the interstellar gas, the magnetic
field and the cosmic rays are comparable, there being some
kind of equipartition of energy among these various components. We recommend the excellent review by \citet{Sofue86} for a comprehensive discussion about 
observational data pertaining to magnetic fields of
spiral galaxies.

\subsection{Parker instability of the interstellar medium}

In a series of papers, Parker presented a study of the dynamical
system comprising the interstellar gas, magnetic
field and cosmic rays in the disk
of a galaxy \citep{Parker66}.  Since the interstellar gas is everywhere
at least partially ionized and has reasonably high electrical
conductivity, the magnetic field remains frozen into it.  The
cosmic ray particles gyrate around this magnetic field and are
confined by it.  As a result, these three components---the 
interstellar gas, magnetic
field and cosmic rays---are coupled to each other and make 
up one unified dynamical system.  Out of these three components,
only the gas remains confined within the gravitational potential
well of the galaxy.  The magnetic field and the cosmic rays are
essentially massless, which try to bulge out of the galactic
disk without being confined by the gravitational field.

\begin{figure}
    \centering
\includegraphics[width= 0.6\textwidth]{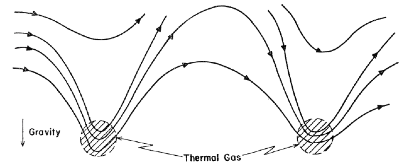}
    \caption{A figure from the paper on Parker instability
    \citep{Parker66}, sketching  how
    the instability arises as a result of the thermal gas flowing down
    the magnetic field lines due to gravity to collect in a few clumps
    in the galactic mid-plane. Reproduced by permission of the AAS.}
\end{figure}

One can think of an equilibrium configuration in which the
magnetic field has the form of straight field lines lying in
the galactic plane.  In the very first paper of the series,
\citet{Parker66} realized that this configuration would be unstable.
This instability is nowadays known as the {\em Parker instability}.
The basic physics of this instability can be elucidated on the
basis of some qualitative arguments without any detailed mathematical
analysis. Suppose the magnetic field has bulged out of the galactic
disk in some region.  
The gas from the upper part of the bulge would
flow due to the gravitational field of the galaxy towards the
galactic disk.  As a result, the upper part of the bulge would
become lighter and more buoyant, triggering an instability and
rising up further.  The gas would keep collecting within the
valleys between the bulges, as shown in Figure~14. Eventually the
rise of the bulge may be halted by the magnetic tension.
It appears from Figure~14 that the resultant configuration may
have clumps of gas along the spiral arm of the galaxy. One
standard way of studying the interstellar gas in external galaxies
is to to use a radio telescope to receive radiation at the
21-cm line of hydrogen.  Radio maps of some external galaxies
show such clumps of gas along spiral arms, like beads on a
string: see, for example, Figure~7 of \cite{Rots75}.  The 
nonlinear evolution and eventual saturation of the
Parker instability have been studied through numerical simulation
by \citet{Mous74}.

\subsection{The galactic dynamo}

If one takes the size of a spiral galaxy as the length scale over which
the galactic magnetic field varies significantly, then the decay
time of the magnetic field turns out to be much larger than the
age of the Universe.  It may appear at first sight that the galactic
magnetic field could therefore be primordial and no mechanism is needed to
sustain it.  However, the disk of a spiral galaxy undergoes differential
rotation, which would be expected to wind up a primordial magnetic
field many times till the relevant length scale becomes much smaller
and the magnetic field is dissipated.  It is clear that a magnetic
field extending along the spiral arms of a galaxy could not be primordial
and we need something like a dynamo mechanism to sustain the magnetic
field in such a configuration.

\citet{Parker71} carried out an analysis of the dynamo problem in
a rectangular slab geometry corresponding to a local region
of the galaxy.  The differential rotation (which was an
uncertain parameter in the solar dynamo problem before
the advent of helioseismology) to be used in the analysis
can easily be obtained if observational data on the rotation
of the galaxy are available. \citet{Parker71} realized that
turbulence in the interstellar
gas in the presence of rotation would be helical and
the $\alpha$-coefficient (Parker did not use the symbol
$\alpha$) associated with it would have opposite signs
above and below the mid-plane of the galaxy.  \citet{Parker71}
set up the $\alpha \Omega$ dynamo equations within the
rectangular slab corresponding to the local region of the
galaxy.  Since the differential rotation stretches out
the magnetic field, the solution suggests a strong
toroidal component of the magnetic field---approximately
in the direction of the spiral arms as seen in the 
observational data.  Parker estimated the growth time
of the dynamo to be of order $10^8$ yr---somewhat smaller
than the age of a typical spiral galaxy.

For readers interested in knowing how our understanding
of the galactic magnetic field evolved in the subsequent
years after the influential work of \citet{Parker71}, we recommend
the excellent review by \citet{Ruz88}.

\def\ub{{\bf u}}

\subsection{Parker limit of magnetic monopoles}

When Parker was selected for
the Henry Norris Russell Lectureship of the American Astronomical Society, he \citep{Parker70}
presented a rather intriguing back-of-the-envelope calculation.
We take the electric field to be zero inside a conductor while
solving an electrostatics problem, because the free electrons inside
the conductor can move around and screen the electric field.  Similarly,
if there were many magnetic monopoles inside a galaxy, they could
have neutralized the magnetic field of the galaxy.  \citet{Parker70} realized 
that the very existence of the galactic magnetic field could be used
for arguing that there could not be too many monopoles in the galaxy.

If there were $n$ magnetic monopoles per unit volume with strength
$g$ moving with velocity $\ub$, then the rate at which the galactic 
magnetic field would perform work per unit volume is $n g \ub. \Bb$.
The energy for doing this work would surely come from the energy
of the galactic magnetic field so that we should have
$$\frac{d}{d t} \left( \frac{B^2}{8 \pi} \right) =
- n g \ub. \Bb.\eqno(13)$$
Note that \citet{Parker70} overlooked the minus sign in his
equation! By demanding that the decay time has to be longer than the
growth time of the magnetic field due to some mechanism (like the
dynamo mechanism), one can put an upper bound on the value of monopole
density $n$. \citet{Parker70} estimated $n < 10^{-26}$ cm$^{-3}$.  This is now
known as the {\em Parker limit}.

\citet{Parker70} presented a short (less than one page) discussion of this topic
in his Russell Lecture clearly to entertain the audience before he
moved into a discussion of more serious stuff.  However, when some
grand unified theories of particle physics suggested the existence
of magnetic monopoles and experiments to detect them were planned in 
the late 1970s and the early 1980s \citep{Cab82}, the Parker limit provided
important guidance in the design of the experiments. Suddenly the
Parker limit became very famous among physicists working in areas
of physics far removed from astrophysics or plasma physics who might  
not know much about Parker's other works.  A more detailed
analysis of the Parker limit was presented by \citet{Parker82}.

\section{Gene as a scientist and as a human being}\label{sec8}

After discussing Gene Parker's science, I
shall now present a
pen portrait of his personality as a scientist
and as a human being.

\subsection{Scientist and member of scientific community}

Since many of Gene Parker's major scientific papers are noted
for the elegance of mathematical analysis, it may seem that the
beauty of mathematical physics might have been what motivated Gene
to do science.  However, he repeatedly told many of us that what
motivated him to a scientific pursuit was his desire to understand
how things work.  When he lived near a railroad yard as a child, he was fascinated to see locomotives move \citep{Parker2014}. He was also
fascinated by cars and aeroplanes and wanted to understand their
basic working principles.  This is certainly not unusual for a 
theoretical physicist, if we think of the example of one of the
greatest theoretical physicists of the twentieth century: Richard 
Feynman, whom Gene admired greatly.  Feynman was also driven by a
desire to understand how things work.

There were, however, big contradictions in Gene's engagement with
technology. It may be expected that somebody who always wanted to
understand how things work would pick up new technology fast and
would be at home in the new world of computers which unfolded
during Gene's career.  That was not the case.  Although Gene always
admitted the importance of numerical simulations and praised those
who were good at it, he himself seemed to be rather afraid of
computers and was unwilling even to touch one for many years.  For
nearly a decade after the rest of the world had switched over to
e-mail, Gene kept sending hand-written letters via air mail.  When
Gene finally started using e-mail, I actually felt saddened that I
would no longer receive those hand-written letters from him.

Since Gene's research had been exclusively based on classical physics,
it may appear that he was temperamentally suited for classical physics.
Again, he told me often that quantum mechanics fascinated him when
he was a student and he often regretted that his creative career
took him along a path in which he never had occasions to use quantum
mechanics.  

Gene was usually a friendly person who was easy to get along. He 
generally had good relations with most of his colleagues and fellow
scientists in the field.  However, sometimes
he could be uncompromising when it came to science. He always
insisted that scientific ideas should be closely and carefully
argued.  He stressed the importance of order-of-magnitude estimates
to check whether an idea worked or not.  He could be very impatient
with ideas which he considered fuzzy and nebulous---especially
when they were put forth by important persons in a pompous manner. Gene
had a famously frosty relationship with Alfv\'en.  He recognized the
importance of Alfv\'en's early contributions to MHD and had also
nominated Alfv\'en for the Nobel Prize in 1964 when he and Chandrasekhar
were invited by the Nobel Committee 
to send nominations.  Gene told me that, although
his relationship with Alfv\'en had already somewhat soured, Chandra
persuaded him to nominate Alfv\'en, arguing that it would be good
for their field if Alfv\'en received the Nobel Prize.  In later years,
as Gene was highly critical of Alfv\'en's newer works, their
relationship nosedived.  Gene was, however, diplomatic enough
not to put his criticisms of Alfv\'en's ideas in his monograph \citep{Parker79a}.
He simply refrained from commenting on those ideas which he 
considered irrelevant.  While talking to students, Gene often
gave the examples of the later Eddington and the later Alfv\'en,
and told us that we should all be careful not to become like them. In a scathing review of
Alfv\'en's book {\em Cosmical Plasmas}, \citet{Cow82}
wrote: ``It was, to say the least, surprising to find a book on cosmical plasmas which did not so much as mention  the work of E. N. Parker.''

Gene was usually kind and encouraging to younger scientists.
However, occasionally he had debates with younger scientists as
well.  As already mentioned in section~5.2, \citet{vabB86} carried
out a simulation of the problem which Gene studied in his classic
1972 paper \citep{Parker72}---how the magnetic field between
two planes gets distorted by footpoint motions on the planes.
Gene thought that the results of the simulation supported his
ideas.  However, Aad van Ballegooijen himself and some others
thought otherwise.  From a distance of nearly four decades, now
this controversy may appear rather irrelevant to us.  However,
it rocked the American solar physics community quite a bit in
the mid-1980s.  When Aad came to give a seminar at the High
Altitude Observatory where I was a postdoc, we found that both
Aad and I were interested in a common scientific question, which we studied
together \citep{vanB88}.  Perhaps I was the only person having regular academic
interactions with both Aad and Gene during the height of their
debate.  Both of them talked to me a few times about this 
controversy.  Although each of them would firmly express his
disagreement with the other, it was a learning experience for me
to see that each referred to the other with extreme respect.
Gene told me that he regarded Aad to be a brilliant young man
and greatly admired his simulations, but was puzzled why Aad was
sticking to what appeared to Gene to be a misjudged interpretation
of the simulation results. I know of only one case when Gene was scathing
and unsparing in his criticism of a younger scientist. 
In a paper \citep{Ionson82} which created a buzz at the time of
its publication but is almost forgotten now, Jim Ionson claimed
that he solved the coronal heating problem by reducing it to the
analogue of an LCR circuit.  Gene felt that the crucial steps in the
derivation of the LCR circuit analogy made no logical sense and
were totally unintelligible. It was the type of paper which Gene
did not want anybody to write.

Gene was always very open about criticisms of his own work and would
readily admit any genuine mistake in his published works if brought to
his attention.  Bernie Roberts, who was a postdoc with Gene in the
1970s, wrote the following to me in an e-mail dated 12 March 2023:

\begin{quote}
    In my first work suggested by Gene (I published it in ApJ in 1976) I found he had made a mistake in an earlier paper. I didn't know how to tell him this but in the end came up with a diplomatic phrase ``this observation means Parker (1974) is redundant''. I asked Gene if he was happy with the comment as he made no remark about it whatever in the draft of my paper. He said he was happy with it. I asked again, could he (ENP) have expressed it better. He thought for a few minutes and then said ``I would say the man is a bloody idiot!''. I have always remembered how he took my correction on the chin, with no excuses offered.
\end{quote}
I visited Chicago a little after I was convinced that the $\alpha \Omega$
dynamo model for the solar cycle proposed by Gene
could not be the final correct model and important modifications were
needed.  I have given an account of my conversation with Gene which
I shall never forget: see p. 182 of \citet{Chou15}.

Gene's self-effacing nature often produced
the opposite of the desired effect.  This
happened for his 60th birthday meeting
organized by some of his colleagues at the
University of Chicago.  After initially
resisting such a meeting, Gene
eventually agreed on the condition that it
would be a low-key meeting for which only a small
number of persons who were close to Gene
should be invited.  However, as the information
about this meeting spread through the community,
receiving an invitation for it became a status
symbol for some senior solar physicists.  I
was a postdoc at the High Altitude Observatory
(HAO) at that time.  One day Peter Gilman working
there stormed into my office and asked me if I was
invited for this meeting.  I told him that
I was invited but would not be able to attend
it, because I had already promised to join a
faculty position in India a few days before the
meeting.  Peter fumed: ``I had worked on some
of the same subjects on which Parker worked 
and I know him personally.  I do not understand
why a senior person like me in this field is
not invited."

I have already mentioned in the Introduction that Gene 
rarely wrote collaborative papers,
most of his papers being single-author.  Presumably, the main
reason behind this is his highly individualistic style
of research.  Using deep physical intuition, he would
think up a mathematically workable model of something that would
capture the essential physics of a complex situation.
He enjoyed working on problems of this kind by himself.
Also, most of his works did not involve the type of lengthy
calculations which he could assign to his students or postdocs
(the typical situation in many theoretical physics
groups around the world). After finishing the work, Gene
would compose his papers with extreme care.  Most of his
papers are models of scientific writing.  Normally we do
not talk about the style of a scientist the way we talk
about the style of an artist like Van Gogh or Cezanne.
But Gene was one rare scientist whose papers can be identified
from the style by somebody who is familiar with his papers.

\subsection{Teacher and supervisor}

Let me say a few words about Gene as teacher.  Gene was an
enthusiastic teacher who taught the first year graduate
level course on electromagnetic theory quite regularly.
I am lucky that, during my first year of graduate school,
he taught a superb course on plasma astrophysics.  It was
rather uncommon for American universities in those days
to offer a
course on plasma astrophysics at the graduate school. I
was quite overwhelmed by the beauty of the subject. 
I was rather undecided till that time as to the area
of physics in which I wanted to pursue my research. It
was this course on plasma astrophysics taught by Gene which made me decide to work in this field.

Gene was a thorough and meticulous teacher, but not flashy
or flamboyant.  He would prepare well for his lectures and
would cover a huge amount of material in each lecture.
His lectures would be very logically structured, with
full derivations worked out on the blackboard in a 
step-by-step manner.  Although his lectures might be a
little dry, a student who had the necessary prerequisite
and followed his classes attentively could get the logical
thread of arguments.  While Gene might not have been the type of
teacher whom students usually think of recommending for
important teaching awards, he was generally regarded as one
of the good teachers in the department whose courses were
extremely useful.

\begin{figure}
    \centering
\includegraphics[width= 0.9\textwidth]{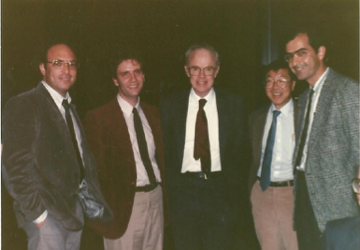}
    \caption{Gene Parker with some of his former PhD
    students at the meeting to celebrate his turning
    60. From left to right: Eugene Levy, Tom
Bogdan, Parker, Boon Chye Low and Kanaris Tsinganos}
\end{figure}

Since Gene was a very individualistic scientist who mostly
worked on his own, one might wonder how he was as a
supervisor.  Gene always enjoyed having young persons
around, with whom he could talk about many things. When
applying for research grants, he would always ask for
funds to support students or postdocs. His group was
never large.  During the four years I worked in his
group, he usually would have two young persons in the
group---either two graduate students or one graduate
student and a postdoc. Gene had 14 graduate students
over his career, whom he had listed in \citet{Parker2014}.
I come towards the end, being his 11th student.  
Gene mentioned that he continued to have regular
ineractions with four of his students over the
years after the completion of their PhD \citep{Parker2014}.
I am lucky to belong to this privileged group along
with B.C.\ Low, Kanaris Tsinganos and Tom Bogdan.
It may
be pointed out that Gene's first student graduated in 1963.
This means that he had no student working with him when
his famous theory of the solar wind was worked out.  

When I first started thinking of working with Gene
after attending his course on plasma astrophysics, some
of the senior astrophysics students cautioned me that,
although Gene was a superb scientist and superb teacher,
he did not have the reputation of being a good research
supervisor.  After I started working with Gene, I realized
that these senior students had a valid point.  It was not
easy to be a student of Gene.  He worked on his own and
wanted the young persons in his group also to work on
their own.  Gene was a very easily approachable person, whom one
could meet virtually any time without an appointment
when he was in his office.  He enjoyed
discussing science with his students. However, apart from
suggestions of very broad and general nature about 
what he considered some of the important unsolved problems
of solar MHD which one could try working on, he never
suggested any specific well-formulated research 
problems---at least to me.  I have to confess that I had
to struggle quite a bit before I could form some idea of
how one selects research problems.  At one stage, when
nothing seemed to work, I thought of quitting research
in astrophysics. Gene persuaded me to continue with his
kind words and encouragement, although he still would
not suggest a definite research problem.  I have given an
account of my experience of working with Gene in my
popular science book: see pp. 122--126 of \citet{Chou15}.

Being very particular about the composition of scientific
papers, Gene would always read the manuscripts of the young
persons in his group very carefully.  When the young person
would get back the manuscript, it would usually be heavily
annotated with suggestions for both science and style.  Gene
regarded this to be the normal duty of a senior scientist and
never expected to be a co-author for such help.  

\subsection{Beyond science}

Gene was a kind and gentle person who always enjoyed hiding his
gentleness underneath an external image of being a no-nonsense
tough guy.  It is not easy to come across such a down-to-earth
person who was so exceptionally free from all kinds of snobbery. His 
assessment of other human beings would always be completely free
from biases of social status, position, race, gender \ldots. Gene
was also a very strong man, who jogged regularly and walked extremely
fast. 
I would usually try to avoid walking with Gene over considerable distances.  Although I was in my twenties when I was a student at Chicago and Gene was in his fifties, I would be panting to keep pace with him.  It would be very embarrassing!  Arieh K\"onigl, who joined as assistant professor at the Universiy of Chicago when I was a student there, wrote to me in an e-mail on 17 March 2022:

\begin{quote}
Regarding Gene's athletic prowess, my own story goes back to my one-day visit to the Department as a faculty candidate, when Gene served as my host. I carried a small suitcase and he just grabbed it and started running up the stairs, with me in toe. I thought to myself at the end of the day that, irrespective of how my job interview would go, I could get back home and brag that, for one whole day when I was in Chicago, Gene Parker had carried my suitcase!

A further note on this story: Even though I only had one day, Gene didn't think he needed to take me to see yet another lab or office; instead, he reckoned (correctly!) that he could impress me even more by taking me to visit the Oriental Institute -- which is what he did.

\end{quote}
The Oriental Institute, later renamed the Institute for the Study of
Ancient Cultures, in a central location in the University of
Chicago campus is an outstanding museum of archaeological specimens
from West Asia and North Africa---mostly discovered by Chicago archaeologists.

Gene took a conscious decision that after retirement in 1995 he would
not pursue his regular astrophysics research and devoted time to his 
other interests.  He would write only a very occasional regular scientific paper when some scientific question occupied his mind.  However, on 
a few occasions, he agreed to write reviews on different aspects of
plasma astrophysics---some of them being of historical nature. He also
wrote the charming little book {\em Conversations on Electric and Magnetic Fields in the Cosmos}
\citep{Parker07}, which shows his sense of humour even when discussing
science.  Gene would spend many pleasurable hours of his retired life
on his other passion: wood carving.  He started on wood carving many
years before his retirement.  He acquired an immense skill for it as
he started spending many hours on wood carving after retirement.  Apart
from a few persons close to Gene, most members of the scientific community
who had admired Gene's science over the years 
would not know about this other aspect
of Gene's interest.  Gene would never make a public display of his
artistic talent and never kept any of his wood carvings in his office.
I saw some of his wood carvings for the first time when I had an
opportunity of visiting his home and was completely amazed by the
professional perfection with which they were executed.
Gene could probably make a living as a professional artist with
his wood carvings if he wished.  A few selected pieces of his wood 
carving are
shown in Figure~16.

\begin{figure}
    \centering
\includegraphics[width= 1.0\textwidth]{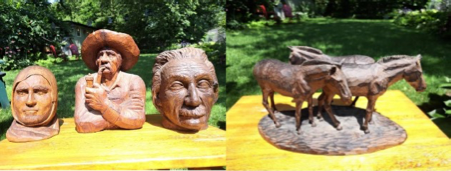}
    \caption{Some of Gene Parker's remarkable wood
    carvings. Credit: Eric Parker.}
\end{figure}

The two great towering figures of theoretical astrophysics at 
Chicago---Chandra and Parker---had one common talent in spite of
the many differences in their personalities.  Both of them were
great story-tellers with amazing memory.  Both could vividly
describe with colourful details some incident which they had
witnessed many years ago.  Gene had a great talent for imitating
other people's speaking styles.  

I realize that I have never talked with Gene about religion and
do not know his views on it.  I always presumed that he did not
have much interest in religious matters.  Gene had extremely
liberal views on politics.  Ronald Reagan was the American
President during the years when I was working  with Gene for
my PhD.  He would often make scathing remarks about Reagan's
policy of interference in the countries of Central America.
He had a tremendous sympathy for people who lived under difficult
circumstances in different parts of the world. The Soviet Union
fascinated him. He had a great respect for the physics research
tradition of that country.  A few years before the breakup
of the Soviet Union (when I was a student at Chicago), Gene
had an opportunity of attending a conference in the Crimean
region and also visited the astrophysics institute at Irkutsk
in the middle of Siberia.  On his return, Gene excitedly described to
me his experience of visiting the Soviet Union and told me some
anecdotes about Zeldovich, whom Gene admired deeply and was looking
forward to meet.

\begin{figure}
    \centering
\includegraphics[width= 1.0\textwidth]{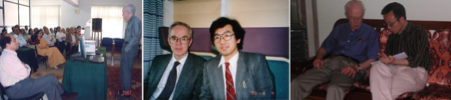}
    \caption{Gene Parker with Asian scientists. (i) Lecturing at the
    Indian Institute of Astrophysics, Bangalore. (ii) With Japanese
    scientist K.\ Shibata. (iii) With P.-F.\ Chen in China.}
\end{figure}
Since this paper is 
written for a journal brought out by the Association of Asia-Pacific
Physics Societies, I end this pen portrait of Gene by mentioning
that he had a special soft corner for the Asia-Pacific region.  Japan
has a long tradition of astrophysics research, including solar physics.
However, when Gene began his scientific career, no other Asian
country had any significant group for solar physics research.
It was during the scientific career of Gene that solar physics
research began in countries like India, China and Korea.  Gene
was a keen observer of these developments and was always willing
to help solar physicists working in these countries.  Although
Gene did not like to travel much, he would never give up an opportunity
of vising an Asian country. Figure~14 shows several
photographs of Gene Parker with Asian scientists. Many 
solar physicists in different Asian
countries told me about Gene's kindnesses and encouragement to their
fledgling scientific communities.  When I decided to return to India
after spending seven years in the USA, academic salaries in India
were typically about one-tenth of the salaries for corresponding
positions in the USA.  Most of my well-wishers in the USA advised
me against this move, which they considered suicidal.  Gene was
one lone person in the USA who stood by my side in that difficult
decision, as described in pp. 133--134 of
\citet{Chou15}.

\section{Concluding remarks}\label{sec9}

We have provided a brief account of Gene Parker's
life and discussed some of his major scientific 
achievements.  In the field of solar MHD, his works
provided the connecting threads among different
aspects of solar activity and transformed the field
into a logically coherent subject.  He also made
fundamental contributions outside solar physics---especially
in our understanding of the galactic magnetic field.
Our discussion of Gene's science should make it clear
why he is regarded as the most impact-making and
tallest figure in the field of plasma astrophysics.
I have also tried to present a pen portrait of Gene
indicating his very unique way of approaching science.

As expected, Gene Parker received many high accolades
during his academic career.  Here is a partial list
of some of the most important academic honours
bestowed on him:
Russell Lectureship (1969); Hale Prize (1978); 
Maxwell Prize (2003); Kyoto Prize (2003); Alfv\'en
Prize (2012); APS
Medal for Exceptional Achievement (2018); Crafoord
Prize (2020).  There are many speculations why he
was not given the Nobel Prize.  As these speculations
are not based on documentary evidence, we refrain from
discussing them.  Since Gene was a self-effacing person
who never blew his own trumpet, during much of his life,
he was not known outside
the community of astrophysicists as much as he should
have been known. However, towards the tail end of his life,
he suddenly came close to becoming a public figure when
he turned out to be the first living person after whom
NASA named an important mission: the Parker Solar Probe!
When this Probe was launched on 12 August 2018, Gene was already
over 91.  Still he travelled to Cape Canaveral with his family
to see the launch of the Probe.  

Gene generally enjoyed
reasonably good health till a couple of years before his
passing away, when deteriorating Parkinson's disease made
it impossible for him to type, which meant that he could
no longer exchange e-mails with his well-wishers.  He spent
the last years of his life in an assisted living facility
near the University of Chicago campus,
where he passed away peacefully, leaving behind Niesje, his
wife for more than 67 years. Niesje passed away on 21 November
2023---some 20 months after Gene's passing away.

Let me end by quoting the message Gene sent on the occasion
of an international Workshop held in Jaipur in connection with
my turning 60.  He wrote:

\begin{quote}
Let me take this festive occasion to congratulate you on a long and 
distinguished research career. I remember your early discussion with 
Chandra on the cultural differences between scientists in the east 
and the west. Chandra was initially vexed with your analysis but soon 
admitted that you had a valid point. Only a great scientist like 
Chandra would recognize the validity of the “upstart” view of an 
“upstart” student. Those were great days and you did not waste any 
time in getting on with your research once you had your degree.

Your treatise THE PHYSICS OF FLUIDS AND PLASMAS has proved to be a 
classic, of which you can be proud. And of which I can be proud that 
you were once my student who then moved on to another book and a long 
distinguished research career. We all salute you.
\end{quote}
Gene was referring to some ideas put forth in \citet{Chou85}  with which Chandra
could not agree. I treasure the above statement from Gene.


\bmhead{Acknowledgments}

I thank Mitsuru Kikuchi for inviting me to write this
review.  The account of Parker's personal life given here
would not have been possible without extensive inputs 
from Eric Parker.  Our thanks go to Susan Kane-Parker for 
organizing and preserving Gene Parker's personal letters
to his parents. I am grateful to Tom Bogdan and Boon
Chye Low for many e-mail exchanges about various aspects of
Parker's science.  I thank Peter Cargill, Marc Kaufman, 
Arieh K\"onigl, 
Eric Priest, Bernie Roberts, Anvar Shukurov and Kanaris
Tsinganos for valuable discussions.  Gopal Hazra and Bibhuti
Kumar Jha helped
me in preparing this manuscript.  
Suggestions from two anonymous referees helped in
improving the manuscript. The Honorary Professorship
offered by the Indian Institute of Science supported my
research.  A large part of this review was written during
my stay at the Max Planck Institute for the History of
Science.  I thank the Alexander von Humboldt Foundation for sponsoring
my visit and thank Alex Blum for stimulating academic
discussions.



\section*{Declarations}

{\bf Conflict of interest} Arnab Rai Choudhuri is an editorial board member for Reviews of Modern Plasma Physics and was not involved in the editorial review or the decision to publish this article. The author declares that there are no other competing interests.


\bibliography{myref}

\begin{thebibliography}{119}
\providecommand{\natexlab}[1]{#1}
\providecommand{\url}[1]{{#1}}
\providecommand{\urlprefix}{URL }
\providecommand{\doi}[1]{\url{https://doi.org/#1}}
\providecommand{\eprint}[2][]{\url{#2}}
 \bibcommenthead

\bibitem[{{Alfv{\'e}n}(1943)}]{Alfven43}
{Alfv{\'e}n} H (1943) {On the Existence of Electromagnetic-Hydrodynamic Waves}.
  Arkiv for Matematik, Astronomi och Fysik 29B:1--7

\bibitem[{{Alfv{\'e}n}(1947)}]{Alf47}
{Alfv{\'e}n} H (1947) {Magneto hydrodynamic waves, and the heating of the solar
  corona}. \mnras 107:211. \doi{10.1093/mnras/107.2.211}

\bibitem[{{Alfv\'en}(1950)}]{Alfven1950}
{Alfv\'en} H (1950) {Cosmical electrodynamics (Clarendon Press, Oxford)}

\bibitem[{{Babcock}(1961)}]{Bab61}
{Babcock} HW (1961) {The Topology of the Sun's Magnetic Field and the 22-YEAR
  Cycle.} \apj 133:572--587. \doi{10.1086/147060}

\bibitem[{{Biermann}(1941)}]{Bier41}
{Biermann} L (1941) {Der gegenw{\"a}rtige Stand der Theorie konvektiver
  Sonnenmodelle}. Vierteljahresschrift der Astronomischen Gesellschaft
  76:194--200

\bibitem[{{Biermann}(1948)}]{Bier48}
{Biermann} L (1948) {{\"U}ber die Ursache der chromosph{\"a}rischen Turbulenz
  und des UV-Exzesses der Sonnenstrahlung}. \zap 25:161

\bibitem[{{Biermann}(1951)}]{Bier51}
{Biermann} L (1951) {Kometenschweife und solare Korpuskularstrahlung}. \zap
  29:274

\bibitem[{Cabrera(1982)}]{Cab82}
Cabrera B (1982) 1st results from a superconductive detector for moving
  magnetic monopoles. \prl 48(20):1378--1381. \doi{10.1103/PhysRevLett.48.1378}

\bibitem[{{Caligari} et~al(1995){Caligari}, {Moreno-Insertis}, and
  {Schussler}}]{cali95}
{Caligari} P, {Moreno-Insertis} F, {Schussler} M (1995) {Emerging Flux Tubes in
  the Solar Convection Zone. I. Asymmetry, Tilt, and Emergence Latitude}. \apj
  441:886. \doi{10.1086/175410}

\bibitem[{{Carrington}(1859)}]{Carring1859}
{Carrington} RC (1859) {Description of a Singular Appearance seen in the Sun on
  September 1, 1859}. \mnras 20:13--15. \doi{10.1093/mnras/20.1.13}

\bibitem[{{Chamberlain}(1960)}]{Chamber60}
{Chamberlain} JW (1960) {Interplanetary Gas.II. Expansion of a Model Solar
  Corona.} \apj 131:47. \doi{10.1086/146805}

\bibitem[{{Chandrasekhar}(1952)}]{Chandra52}
{Chandrasekhar} S (1952) On the inhibition of convection by a magnetic field.
  Philosophical Magazine 43(340):501--532. \doi{10.1080/14786440508520205},
  \urlprefix\url{http://dx.doi.org/10.1080/14786440508520205},
  {\href{https://arxiv.org/abs/http://dx.doi.org/10.1080/14786440508520205}{{https://arxiv.org/abs/http://dx.doi.org/10.1080/14786440508520205}}}

\bibitem[{{Chandrasekhar} and {Fermi}(1953)}]{Chandra53}
{Chandrasekhar} S, {Fermi} E (1953) {Magnetic Fields in Spiral Arms.} \apj
  118:113. \doi{10.1086/145731}

\bibitem[{{Chapman}(1957)}]{Chap57}
{Chapman} S (1957) {Notes on the solar corona and the terrestrial ionosphere}.
  Smithsonian Contributions to Astrophysics 2:1--14

\bibitem[{{Charbonneau}(2010)}]{Charbonneau10}
{Charbonneau} P (2010) {Dynamo Models of the Solar Cycle}. Living Rev Solar
  Phys 7:3. \doi{10.12942/lrsp-2010-3}

\bibitem[{{Charbonneau}(2014)}]{Charbonneau14}
{Charbonneau} P (2014) {Solar Dynamo Theory}. \araa 52:251--290.
  \doi{10.1146/annurev-astro-081913-040012}

\bibitem[{{Choudhuri}(1985)}]{Chou85}
{Choudhuri} AR (1985) {Practising Western science outside the West: Personal
  observations on the Indian scene}. Social Studies of Science 15:475--505

\bibitem[{{Choudhuri}(1989)}]{Chou89}
{Choudhuri} AR (1989) {The evolution of loop structures in flux rings within
  the solar convection zone}. \solphys 123:217--239. \doi{10.1007/BF00149104}

\bibitem[{{Choudhuri}(1998)}]{Chou98}
{Choudhuri} AR (1998) {The physics of fluids and plasmas : an introduction for
  astrophysicists (Cambridge: Cambridge University Press)}

\bibitem[{{Choudhuri}(2010)}]{Chou2010}
{Choudhuri} AR (2010) {Astrophysics for Physicists (Cambridge University
  Press)}

\bibitem[{{Choudhuri}(2011)}]{Chou11}
{Choudhuri} AR (2011) {The origin of the solar magnetic cycle}. Pramana
  77:77--96. \doi{10.1007/s12043-011-0113-4},
  {\href{https://arxiv.org/abs/1103.3385}{{https://arxiv.org/abs/arXiv:1103.3385}}}
  {[astro-ph.SR]}

\bibitem[{{Choudhuri}(2014)}]{Chou13}
{Choudhuri} AR (2014) {The irregularities of the sunspot cycle and their
  theoretical modelling}. Indian Journal of Physics 88:877--884.
  \doi{10.1007/s12648-014-0481-y},
  {\href{https://arxiv.org/abs/1312.3408}{{https://arxiv.org/abs/arXiv:1312.3408}}}
  {[astro-ph.SR]}

\bibitem[{{Choudhuri}(2015)}]{Chou15}
{Choudhuri} AR (2015) {Nature's third cycle: a story of sunspots (Oxford
  University Press)}. \doi{10.1093/acprof:oso/9780199674756.001.0001}

\bibitem[{{Choudhuri}(2017)}]{Chou17}
{Choudhuri} AR (2017) {Starspots, stellar cycles and stellar flares: Lessons
  from solar dynamo models}. Science China Physics, Mechanics, and Astronomy
  60(1):19601. \doi{10.1007/s11433-016-0413-7},
  {\href{https://arxiv.org/abs/1612.02544}{{https://arxiv.org/abs/arXiv:1612.02544}}}
  {[astro-ph.SR]}

\bibitem[{{Choudhuri}(2023)}]{Chou23}
{Choudhuri} AR (2023) {The emergence and growth of the flux transport dynamo
  model of the sunspot cycle}. Reviews of Modern Plasma Physics 7(1):18.
  \doi{10.1007/s41614-023-00120-9},
  {\href{https://arxiv.org/abs/2212.14617}{{https://arxiv.org/abs/arXiv:2212.14617}}}
  {[astro-ph.SR]}

\bibitem[{{Choudhuri} et~al(1995){Choudhuri}, {Sch\"ussler}, and
  {Dikpati}}]{CSD95}
{Choudhuri} AR, {Sch\"ussler} M, {Dikpati} M (1995) {The solar dynamo with
  meridional circulation.} \aap 303:L29--L32

\bibitem[{{Choudhuri} et~al(2007){Choudhuri}, {Chatterjee}, and
  {Jiang}}]{CCJ07}
{Choudhuri} AR, {Chatterjee} P, {Jiang} J (2007) {Predicting Solar Cycle 24
  With a Solar Dynamo Model}. Physical Review Letters 98:131103.
  \doi{10.1103/PhysRevLett.98.131103},
  {\href{https://arxiv.org/abs/astro-ph/0701527}{{https://arxiv.org/abs/astro-ph/0701527}}}

\bibitem[{{Cowling}(1933)}]{Cowling33}
{Cowling} TG (1933) {The magnetic field of sunspots}. \mnras 94:39--48.
  \doi{10.1093/mnras/94.1.39}

\bibitem[{{Cowling}(1953)}]{Cowling53}
{Cowling} TG (1953) {Solar Electrodynamics}. In: {Kuiper} GP (ed) The Sun. p
  532

\bibitem[{{Cowling}(1982)}]{Cow82}
{Cowling} TG (1982) {A review of: ``Cosmical plasmas''}. Geophysical and
  Astrophysical Fluid Dynamics 21(3):324--327. \doi{10.1080/03091928208209024}

\bibitem[{{Davis} and {Greenstein}(1951)}]{Davis51}
{Davis} JLeverett, {Greenstein} JL (1951) {The Polarization of Starlight by
  Aligned Dust Grains.} \apj 114:206. \doi{10.1086/145464}

\bibitem[{{D'Silva} and {Choudhuri}(1993)}]{Dsilva93}
{D'Silva} S, {Choudhuri} AR (1993) {A theoretical model for tilts of bipolar
  magnetic regions}. \aap 272:621--633

\bibitem[{Dungey(1953)}]{Dungey53}
Dungey J (1953) Conditions for the occurrence of electrical discharges in
  astrophysical systems. Philosophical Magazine 44(354):725--738

\bibitem[{{Durney}(1995)}]{Durney95}
{Durney} BR (1995) {On a Babcock-Leighton dynamo model with a deep-seated
  generating layer for the toroidal magnetic field}. \solphys 160:213--235.
  \doi{10.1007/BF00732805}

\bibitem[{{Edl{\'e}n}(1943)}]{Edlen43}
{Edl{\'e}n} B (1943) {Die Deutung der Emissionslinien im Spektrum der
  Sonnenkorona. Mit 6 Abbildungen.} \zap 22:30

\bibitem[{Elsasser(1946)}]{Elsasser46}
Elsasser W (1946) Induction effects in terrestrial magnetism .1. theory.
  PHYSICAL REVIEW 69(3-4):106--116. \doi{10.1103/PhysRev.69.106}

\bibitem[{{Fan}(2021)}]{Fan21}
{Fan} Y (2021) {Magnetic fields in the solar convection zone}. Living Reviews
  in Solar Physics 18(1):5. \doi{10.1007/s41116-021-00031-2}

\bibitem[{{Fan} et~al(1993){Fan}, {Fisher}, and {Deluca}}]{Fan93}
{Fan} Y, {Fisher} GH, {Deluca} EE (1993) {The Origin of Morphological
  Asymmetries in Bipolar Active Regions}. \apj 405:390. \doi{10.1086/172370}

\bibitem[{{Fermi}(1949)}]{Fermi49}
{Fermi} E (1949) {On the Origin of the Cosmic Radiation}. Physical Review
  75(8):1169--1174. \doi{10.1103/PhysRev.75.1169}

\bibitem[{{Ferraro}(1937)}]{Ferraro37}
{Ferraro} VCA (1937) {The non-uniform rotation of the Sun and its magnetic
  field}. \mnras 97:458. \doi{10.1093/mnras/97.6.458}

\bibitem[{{Forbush}(1954)}]{Forbush54}
{Forbush} SE (1954) {World-Wide Cosmic-Ray Variations, 1937-1952}. \jgr
  59(4):525--542. \doi{10.1029/JZ059i004p00525}

\bibitem[{{Glatzmaier} and {Roberts}(1995)}]{Glatz95}
{Glatzmaier} GA, {Roberts} PH (1995) {A three-dimensional self-consistent
  computer simulation of a geomagnetic field reversal}. \nat
  377(6546):203--209. \doi{10.1038/377203a0}

\bibitem[{{Gringauz} et~al(1960){Gringauz}, {Bezrokikh}, {Ozerov}, and
  {Rybchinskii}}]{Grin60}
{Gringauz} KI, {Bezrokikh} VV, {Ozerov} VD, et~al (1960) {A Study of the
  Interplanetary Ionized Gas, High-Energy Electrons and Corpuscular Radiation
  from the Sun by Means of the Three-Electrode Trap for Charged Particles on
  the Second Soviet Cosmic Rocket}. Soviet Physics Doklady 5:361

\bibitem[{{Hale}(1908)}]{Hale1909}
{Hale} GE (1908) {On the Probable Existence of a Magnetic Field in Sun-Spots}.
  \apj 28:315. \doi{10.1086/141602}

\bibitem[{{Hale} et~al(1919){Hale}, {Ellerman}, {Nicholson}, and
  {Joy}}]{Hale19}
{Hale} GE, {Ellerman} F, {Nicholson} SB, et~al (1919) {The Magnetic Polarity of
  Sun-Spots}. \apj 49:153. \doi{10.1086/142452}

\bibitem[{{Hasan}(1985)}]{Hasan85}
{Hasan} SS (1985) {Convective instability in a solar flux tube. II - Nonlinear
  calculations with horizontal radiative heat transport and finite viscosity}.
  \aap 143(1):39--45

\bibitem[{Hess(1912)}]{Hess12}
Hess V (1912) Observations in low level radiation during seven free balloon
  flights. Physikalische Zeitschrift 13:1084--1091

\bibitem[{{Hiltner}(1949)}]{Hilt49}
{Hiltner} WA (1949) {Polarization of Light from Distant Stars by Interstellar
  Medium}. Science 109(2825):165. \doi{10.1126/science.109.2825.165}

\bibitem[{{Ionson}(1982)}]{Ionson82}
{Ionson} JA (1982) {Resonant electrodynamic heating of stellar coronal loops -
  an LRC circuit analog}. \apj 254:318--334. \doi{10.1086/159736}

\bibitem[{{Jokipii}(1971)}]{Jokipii71}
{Jokipii} JR (1971) {Propagation of cosmic rays in the solar wind.} Reviews of
  Geophysics and Space Physics 9:27--87. \doi{10.1029/RG009i001p00027}

\bibitem[{{Jokipii} and {Parker}(1969)}]{Parker69}
{Jokipii} JR, {Parker} EN (1969) {Stochastic Aspects of Magnetic Lines of Force
  with Application to Cosmic-Ray Propagation}. \apj 155:777.
  \doi{10.1086/149909}

\bibitem[{{Kiepenheuer}(1953)}]{Kiepen53}
{Kiepenheuer} KO (1953) {Solar Activity}. In: {Kuiper} GP (ed) The Sun. p 322

\bibitem[{{Krause}(1993)}]{Krause93}
{Krause} F (1993) {The Cosmic Dynamo: From $t= -\infty$ to Cowling's Theorem. A
  Review on History}. In: {Krause} F, {Radler} KH, {Rudiger} G (eds) The Cosmic
  Dynamo: IAU Symposium no. 157, p 487

\bibitem[{{Kuiper}(1953)}]{Kuiper53}
{Kuiper} GP (1953) {The Sun (The University of Chicago Press)}

\bibitem[{{Kuperus} et~al(1981){Kuperus}, {Ionson}, and {Spicer}}]{Kuperus81}
{Kuperus} M, {Ionson} JA, {Spicer} DS (1981) {On the theory of coronal heating
  mechanisms}. \araa 19:7--40. \doi{10.1146/annurev.aa.19.090181.000255}

\bibitem[{{Leer} et~al(1982){Leer}, {Holzer}, and {Fla}}]{Leer82}
{Leer} E, {Holzer} TE, {Fla} T (1982) {Acceleration of the solar wind.} \ssr
  33(1-2):161--200. \doi{10.1007/BF00213253}

\bibitem[{{Leighton}(1969)}]{Leighton69}
{Leighton} RB (1969) {A Magneto-Kinematic Model of the Solar Cycle}. \apj
  156:1--26. \doi{10.1086/149943}

\bibitem[{Low(2023)}]{Low23}
Low BC (2023) Topological nature of the parker magnetostatic theorem. Physics
  of Plasmas 30(1). \doi{10.1063/5.0124164}

\bibitem[{{Lyot}(1939)}]{Lyot39}
{Lyot} B (1939) {The study of the solar corona and prominences without eclipses
  (George Darwin Lecture, 1939)}. \mnras 99:580. \doi{10.1093/mnras/99.8.580}

\bibitem[{{Mestel}(1968)}]{Mestel68}
{Mestel} L (1968) {Magnetic braking by a stellar wind-I}. \mnras 138:359.
  \doi{10.1093/mnras/138.3.359}

\bibitem[{{Mestel}(1999)}]{Mestel99}
{Mestel} L (1999) {Stellar magnetism (Clarendon Press, Oxford)}

\bibitem[{{Moffatt}(1978)}]{Moff78}
{Moffatt} HK (1978) {Magnetic field generation in electrically conducting
  fluids (Cambridge University Press)}

\bibitem[{{Moreno-Insertis}(1986)}]{Moreno86}
{Moreno-Insertis} F (1986) {Nonlinear time-evolution of kink-unstable magnetic
  flux tubes in the convective zone of the sun}. \aap 166(1-2):291--305

\bibitem[{{Mouschovias}(1974)}]{Mous74}
{Mouschovias} TC (1974) {Static Equilibria of the Interstellar Gas in the
  Presence of Magnetic and Gravitational Fields: Large-Scale Condensations}.
  \apj 192:37--50. \doi{10.1086/153032}

\bibitem[{{Obridko} and {Vaisberg}(2017)}]{Obridko17}
{Obridko} VN, {Vaisberg} OL (2017) {On the history of the solar wind
  discovery}. Solar System Research 51(2):165--169.
  \doi{10.1134/S0038094617020058}

\bibitem[{{Pallavicini} et~al(1981){Pallavicini}, {Golub}, {Rosner}, {Vaiana},
  {Ayres}, and {Linsky}}]{Pall81}
{Pallavicini} R, {Golub} L, {Rosner} R, et~al (1981) {Relations among stellar
  X-ray emission observed from Einstein, stellar rotation and bolometric
  luminosity.} \apj 248:279--290. \doi{10.1086/159152}

\bibitem[{{Parker}(1955{\natexlab{a}})}]{Parker55b}
{Parker} EN (1955{\natexlab{a}}) {Hydromagnetic Dynamo Models.} \apj
  122:293--314. \doi{10.1086/146087}

\bibitem[{{Parker}(1955{\natexlab{b}})}]{Parker55a}
{Parker} EN (1955{\natexlab{b}}) {The Formation of Sunspots from the Solar
  Toroidal Field.} \apj 121:491. \doi{10.1086/146010}

\bibitem[{{Parker}(1957)}]{Parker57}
{Parker} EN (1957) {Sweet's Mechanism for Merging Magnetic Fields in Conducting
  Fluids}. \jgr 62(4):509--520. \doi{10.1029/JZ062i004p00509}

\bibitem[{{Parker}(1958)}]{Parker58}
{Parker} EN (1958) {Dynamics of the Interplanetary Gas and Magnetic Fields.}
  \apj 128:664. \doi{10.1086/146579}

\bibitem[{{Parker}(1963{\natexlab{a}})}]{Parker63a}
{Parker} EN (1963{\natexlab{a}}) {Interplanetary dynamical processes
  (Interscience Publishers, New York)}

\bibitem[{{Parker}(1963{\natexlab{b}})}]{Parker63b}
{Parker} EN (1963{\natexlab{b}}) {Kinematical Hydromagnetic Theory and its
  Application to the Low Solar Photosphere.} \apj 138:552. \doi{10.1086/147663}

\bibitem[{{Parker}(1964)}]{Parker64}
{Parker} EN (1964) {The Solar Wind}. Scientific American 210.4:66--76

\bibitem[{{Parker}(1965{\natexlab{a}})}]{Parker1965}
{Parker} EN (1965{\natexlab{a}}) {Dynamical Theory of the Solar Wind}. \ssr
  4(5-6):666--708. \doi{10.1007/BF00216273}

\bibitem[{{Parker}(1965{\natexlab{b}})}]{Parker65}
{Parker} EN (1965{\natexlab{b}}) {The passage of energetic charged particles
  through interplanetary space}. Planetary and Space Science 13(1):9--49.
  \doi{10.1016/0032-0633(65)90131-5}

\bibitem[{{Parker}(1966)}]{Parker66}
{Parker} EN (1966) {The Dynamical State of the Interstellar Gas and Field}.
  \apj 145:811. \doi{10.1086/148828}

\bibitem[{{Parker}(1970)}]{Parker70}
{Parker} EN (1970) {The Origin of Magnetic Fields}. \apj 160:383.
  \doi{10.1086/150442}

\bibitem[{{Parker}(1971)}]{Parker71}
{Parker} EN (1971) {The Generation of Magnetic Fields in Astrophysical Bodies.
  II. The Galactic Field}. \apj 163:255. \doi{10.1086/150765}

\bibitem[{{Parker}(1972)}]{Parker72}
{Parker} EN (1972) {Topological Dissipation and the Small-Scale Fields in
  Turbulent Gases}. \apj 174:499. \doi{10.1086/151512}

\bibitem[{{Parker}(1975)}]{Parker75}
{Parker} EN (1975) {The generation of magnetic fields in astrophysical bodies.
  X - Magnetic buoyancy and the solar dynamo}. \apj 198:205--209.
  \doi{10.1086/153593}

\bibitem[{{Parker}(1978)}]{Parker78}
{Parker} EN (1978) {Hydraulic concentration of magnetic fields in the solar
  photosphere. VI. Adiabatic cooling and concentration in downdrafts.} \apj
  221:368--377. \doi{10.1086/156035}

\bibitem[{{Parker}(1979{\natexlab{a}})}]{Parker79a}
{Parker} EN (1979{\natexlab{a}}) {Cosmical magnetic fields: Their origin and
  their activity (Clarendon Press, Oxford)}

\bibitem[{{Parker}(1979{\natexlab{b}})}]{Parker79b}
{Parker} EN (1979{\natexlab{b}}) {Sunspots and the physics of magnetic flux
  tubes. I. The general nature of the sunspots.} \apj 230:905--923.
  \doi{10.1086/157150}

\bibitem[{{Parker}(1983)}]{Parker83}
{Parker} EN (1983) {Magnetic Neutral Sheets in Evolving Fields - Part Two -
  Formation of the Solar Corona}. \apj 264:642. \doi{10.1086/160637}

\bibitem[{{Parker}(1988)}]{Parker88}
{Parker} EN (1988) {Nanoflares and the Solar X-Ray Corona}. \apj 330:474.
  \doi{10.1086/166485}

\bibitem[{{Parker}(1989)}]{Parker89}
{Parker} EN (1989) {Tangential discontinuities and the optical analogy for
  stationary fields II. The optical analogy}. Geophysical and Astrophysical
  Fluid Dynamics 45(3):169--182. \doi{10.1080/03091928908208898}

\bibitem[{{Parker}(1993)}]{Parker93}
{Parker} EN (1993) {A Solar Dynamo Surface Wave at the Interface between
  Convection and Nonuniform Rotation}. \apj 408:707. \doi{10.1086/172631}

\bibitem[{{Parker}(1994)}]{Parker94}
{Parker} EN (1994) {Spontaneous current sheets in magnetic fields : with
  applications to stellar x-rays (Oxford University Press)}

\bibitem[{{Parker}(1996)}]{Parker96}
{Parker} EN (1996) {S. Chandrasekhar and Magnetohydrodynamics}. Journal of
  Astrophysics and Astronomy 17(3-4):147--166. \doi{10.1007/BF02702301}

\bibitem[{{Parker}(2007)}]{Parker07}
{Parker} EN (2007) {Conversations on Electric and Magnetic Fields in the Cosmos
  (Princeton University Press)}

\bibitem[{{Parker}(2014)}]{Parker2014}
{Parker} EN (2014) {Reminiscing my sixty year pursuit of the physics of the Sun
  and the Galaxy}. Research in Astronomy and Astrophysics 14(1):1-14.
  \doi{10.1088/1674-4527/14/1/001}

\bibitem[{{Petschek}(1964)}]{Pets64}
{Petschek} HE (1964) {Magnetic Field Annihilation}. In: NASA Special
  Publication, vol~50. p 425

\bibitem[{{Pneuman} and {Kopp}(1971)}]{Pneu71}
{Pneuman} GW, {Kopp} RA (1971) {Gas-Magnetic Field Interactions in the Solar
  Corona}. \solphys 18(2):258--270. \doi{10.1007/BF00145940}

\bibitem[{{Pontin} and {Priest}(2022)}]{Pontin22}
{Pontin} DI, {Priest} ER (2022) {Magnetic reconnection: MHD theory and
  modelling}. Living Reviews in Solar Physics 19(1):1.
  \doi{10.1007/s41116-022-00032-9}

\bibitem[{{Priest}(2014)}]{Priest14}
{Priest} E (2014) {Magnetohydrodynamics of the Sun (Cambridge University
  Press)}. \doi{10.1017/CBO9781139020732}

\bibitem[{{Priest} and {Forbes}(2000)}]{Priest00}
{Priest} E, {Forbes} T (2000) {Magnetic Reconnection (Cambridge University
  Press)}

\bibitem[{{Priest}(1982)}]{Priest82}
{Priest} ER (1982) {Solar magnetohydrodynamics (D. Reidel, Dordrecht)}

\bibitem[{{Raouafi} et~al(2023){Raouafi}, {Matteini}, {Squire}, {Badman},
  {Velli}, {Klein}, {Chen}, {Matthaeus}, {Szabo}, {Linton}, {Allen}, {Szalay},
  {Bruno}, {Decker}, {Akhavan-Tafti}, {Agapitov}, {Bale}, {Bandyopadhyay},
  {Battams}, {Ber{\v{c}}i{\v{c}}}, {Bourouaine}, {Bowen}, {Cattell},
  {Chandran}, {Chhiber}, {Cohen}, {D'Amicis}, {Giacalone}, {Hess}, {Howard},
  {Horbury}, {Jagarlamudi}, {Joyce}, {Kasper}, {Kinnison}, {Laker}, {Liewer},
  {Malaspina}, {Mann}, {McComas}, {Niembro-Hernandez}, {Nieves-Chinchilla},
  {Panasenco}, {Pokorn{\'y}}, {Pusack}, {Pulupa}, {Perez}, {Riley},
  {Rouillard}, {Shi}, {Stenborg}, {Tenerani}, {Verniero}, {Viall}, {Vourlidas},
  {Wood}, {Woodham}, and {Woolley}}]{Raou23}
{Raouafi} NE, {Matteini} L, {Squire} J, et~al (2023) {Parker Solar Probe: Four
  Years of Discoveries at Solar Cycle Minimum}. \ssr 219(1):8.
  \doi{10.1007/s11214-023-00952-4},
  {\href{https://arxiv.org/abs/2301.02727}{{https://arxiv.org/abs/arXiv:2301.02727}}}
  {[astro-ph.SR]}

\bibitem[{{Rots}(1975)}]{Rots75}
{Rots} AH (1975) {Distribution and kinematics of neutral hydrogen in the spiral
  galaxy M81. II. Analysis.} \aap 45:43--55

\bibitem[{{Ruzmaikin} et~al(1988){Ruzmaikin}, {Sokolov}, and
  {Shukurov}}]{Ruz88}
{Ruzmaikin} A, {Sokolov} D, {Shukurov} A (1988) {Magnetism of spiral galaxies}.
  \nat 336(6197):341--347. \doi{10.1038/336341a0}

\bibitem[{{Sakurai}(1985)}]{Saku85}
{Sakurai} T (1985) {Magnetic stellar winds: a 2-D generalization of the
  Weber-Davis model.} \aap 152:121--129

\bibitem[{{Schatzman}(1962)}]{Schatz62}
{Schatzman} E (1962) {A theory of the role of magnetic activity during star
  formation}. Annales d'Astrophysique 25:18

\bibitem[{{Schwarzschild}(1948)}]{Schwarz48}
{Schwarzschild} M (1948) {On Noise Arising from the Solar Granulation.} \apj
  107:1. \doi{10.1086/144983}

\bibitem[{{Shibata} and {Magara}(2011)}]{Shibata11}
{Shibata} K, {Magara} T (2011) {Solar Flares: Magnetohydrodynamic Processes}.
  Living Reviews in Solar Physics 8(1):6. \doi{10.12942/lrsp-2011-6}

\bibitem[{{Sofue} et~al(1986){Sofue}, {Fujimoto}, and {Wielebinski}}]{Sofue86}
{Sofue} Y, {Fujimoto} M, {Wielebinski} R (1986) {Global structure of magnetic
  fields in spiral galaxies.} \araa 24:459--497.
  \doi{10.1146/annurev.aa.24.090186.002331}

\bibitem[{{Spruit}(1979)}]{Spruit79}
{Spruit} HC (1979) {Convective collapse of flux tubes.} \solphys
  61(2):363--378. \doi{10.1007/BF00150420}

\bibitem[{{Spruit}(1981)}]{Spr81}
{Spruit} HC (1981) {Motion of magnetic flux tubes in the solar convection zone
  and chromosphere.} \aap 98:155--160

\bibitem[{{Steenbeck} and {Krause}(1969)}]{SK69}
{Steenbeck} M, {Krause} F (1969) {Zur Dynamotheorie stellarer und planetarer
  Magnetfelder I. Berechnung sonnen\"ahnlicher Wechselfeldgeneratoren}.
  Astronomische Nachrichten 291:49--84. \doi{10.1002/asna.19692910201}

\bibitem[{{Steenbeck} et~al(1966){Steenbeck}, {Krause}, and
  {R{\"a}dler}}]{Steen66}
{Steenbeck} M, {Krause} F, {R{\"a}dler} KH (1966) {Berechnung der mittleren
  Lorentz-Feldst{\"a}rke v X B f{\"u}r ein elektrisch leitendes Medium in
  turbulenter, durch Coriolis-Kr{\"a}fte beeinflu{\ss}ter Bewegung}.
  Zeitschrift Naturforschung Teil A 21:369. \doi{10.1515/zna-1966-0401}

\bibitem[{{Stenflo}(1973)}]{Sten73}
{Stenflo} JO (1973) {Magnetic-Field Structure of the Photospheric Network}.
  \solphys 32(1):41--63. \doi{10.1007/BF00152728}

\bibitem[{{Sweet}(1958)}]{Sweet58}
{Sweet} PA (1958) {The Neutral Point Theory of Solar Flares}. In: {Lehnert} B
  (ed) Electromagnetic Phenomena in Cosmical Physics: IAU Symposium no. 6, p
  123

\bibitem[{{Turner} et~al(1982){Turner}, {Parker}, and {Bogdan}}]{Parker82}
{Turner} MS, {Parker} EN, {Bogdan} TJ (1982) {Magnetic monopoles and the
  survival of galactic magnetic fields}. Physical Review D 26(6):1296--1305.
  \doi{10.1103/PhysRevD.26.1296}

\bibitem[{{van Ballegooijen}(1986)}]{vabB86}
{van Ballegooijen} AA (1986) {Cascade of Magnetic Energy as a Mechanism of
  Coronal Heating}. \apj 311:1001. \doi{10.1086/164837}

\bibitem[{{van Ballegooijen} and {Choudhuri}(1988)}]{vanB88}
{van Ballegooijen} AA, {Choudhuri} AR (1988) {The Possible Role of Meridional
  Flows in Suppressing Magnetic Buoyancy}. \apj 333:965. \doi{10.1086/166805}

\bibitem[{{Wang} et~al(1991){Wang}, {Sheeley}, and {Nash}}]{WSN91}
{Wang} YM, {Sheeley} NRJr., {Nash} AG (1991) {A new solar cycle model including
  meridional circulation}. \apj 383:431--442. \doi{10.1086/170800}

\bibitem[{{Weber} and {Davis}(1967)}]{WD67}
{Weber} EJ, {Davis} JLeverett (1967) {The Angular Momentum of the Solar Wind}.
  \apj 148:217--227. \doi{10.1086/149138}

\bibitem[{{Weiss}(1966)}]{Weiss66}
{Weiss} NO (1966) {The Expulsion of Magnetic Flux by Eddies}. Proceedings of
  the Royal Society of London Series A 293(1434):310--328.
  \doi{10.1098/rspa.1966.0173}

\bibitem[{{Yoshimura}(1975)}]{Yosh75}
{Yoshimura} H (1975) {Solar-cycle dynamo wave propagation.} \apj 201:740--748.
  \doi{10.1086/153940}

\bibitem[{{Zwaan}(1985)}]{Zwaan85}
{Zwaan} C (1985) {The Emergence of Magnetic Flux}. \solphys 100:397.
  \doi{10.1007/BF00158438}

\end{thebibliography}

\end{document}